\documentclass[aps,prd,twocolumn,preprintnumbers]{revtex4-1}
\usepackage{amsmath}
\usepackage{amsfonts}
\usepackage{graphicx}
\usepackage[config=altsf]{subfig}
\usepackage{feynmp}
\usepackage{dsfont}
\usepackage{array}
\usepackage{slashed}
\usepackage{afterpage}
\usepackage{appendix}
\usepackage{multirow}
\usepackage{psfrag}
\def\lesssim{\mathrel{\mathpalette\vereq<}}
\def\gtrsim{\mathrel{\mathpalette\vereq>}}

\begin{document}
\title{Flavored Dark Matter, and Its Implications for Direct Detection
and Colliders}

\author{Prateek Agrawal}
\affiliation{Department of Physics, University of Maryland,
College Park, MD 20742}
\author{Steve Blanchet}
\affiliation{Instituto de F\'isica Te\'orica,
IFT-UAM/CSIC Nicolas Cabrera 15,
UAM Cantoblanco, 28049 Madrid, Spain
}
\author{Zackaria Chacko}
\affiliation{Department of Physics, University of Maryland,
College Park, MD 20742}
\author{Can Kilic}
\affiliation{Department of Physics and Astronomy, Rutgers University,
Piscataway NJ 08854}
\affiliation{Theory Group, Department of Physics and Texas Cosmology Center,
The University of Texas at Austin,
Austin, TX 78712}


\begin{abstract}

We consider theories where the dark matter particle carries flavor
quantum numbers, and has renormalizable contact interactions with the
Standard Model fields. The phenomenology of this scenario depends
sensitively on whether dark matter carries lepton flavor, quark flavor
or its own internal flavor quantum numbers. We show that each of these
possibilities is associated with a characteristic type of vertex, has
different implications for direct detection experiments and gives rise
to distinct collider signatures. We find that the region of parameter
space where dark matter has the right abundance to be a thermal relic is
in general within reach of current direct detection experiments. We
focus on a class of models where dark matter carries tau flavor, and
show that the collider signals of these models include events with four
or more isolated leptons and missing energy. A full simulation of the
signal and backgrounds, including detector effects, shows that in a
significant part of parameter space these theories can be discovered
above Standard Model backgrounds at the Large Hadron Collider. We also
study the extent to which flavor and charge correlations among the final
state leptons allows models of this type to be distinguished from
theories where dark matter couples to leptons but does not carry flavor.

\end{abstract}
\maketitle
\preprint{UMD-PP-011-014}
\preprint{RUNHETC-2011-17}
\preprint{UTTG-19-11}
\preprint{TCC-020-11}

\section{Introduction}

It is now well established that about 80\% of the matter in the universe
is in fact dark matter, rather than visible matter
\cite{Komatsu:2010fb}. However, the masses
and interactions of the particles of which dark matter is composed are
not known. One simple and well-motivated possibility is that dark matter
is made up of particles with masses close to the weak scale that have
weak scale annihilation cross section to Standard Model (SM)
particles. Dark matter candidates with these properties neatly fit
into the `Weakly Interacting Massive Particle' (WIMP) paradigm, and
therefore naturally tend to have the
right relic abundance to explain observations.

The matter fields ($Q, U^c, D^c, L, E^c$) of the SM each come in three
copies, or flavors, that differ only in their masses. This reflects the
fact that the Lagrangian of the SM possesses an approximate U(3$)^5$
flavor symmetry acting on the matter fields, which is explicitly broken
by the Yukawa couplings that generate the quark and lepton masses. An
interesting possibility is that the dark matter field, which we label by
$\chi$, also carries flavor quantum numbers, with the physical dark
matter particle being the lightest of three copies. Several specific
dark matter candidates of this type have been studied extensively in the
literature, including sneutrino dark matter \cite{Ibanez:1983kw,
Ellis:1983ew, Hagelin:1984wv, Goodman:1984dc, Freese:1985qw,
Falk:1994es} (for recent work see~\cite{MarchRussell:2009aq}) in the
Minimal Supersymmetric Standard Model (MSSM), and Kaluza-Klein (KK)
neutrino dark matter \cite{Servant:2002aq} in models with a universal
extra dimension (UED).
Other realizations of flavored dark matter that have received recent
study include theories where dark matter couples primarily to quarks,
potentially giving rise to interesting flavor violating
signals~\cite{Kile:2011mn,Kamenik:2011nb}. It has also been shown that
flavored dark matter may play a role in explaining the baryon
asymmetry~\cite{Cui:2011qe}, and that extending the SM flavor structure
to the dark sector can explain the stability of dark
matter~\cite{Batell:2011tc}.

In this paper we first consider the general properties of theories where
dark matter carries flavor quantum numbers and has renormalizable
contact interactions with the SM fields. We classify the different
models of this type, consider how the SM flavor structure can be
extended to incorporate dark matter, and study the implications of these
theories for direct detection and collider experiments. We find that the
phenomenology depends sensitively on whether dark matter carries lepton
flavor, quark flavor or internal flavor quantum numbers.  Most models of
flavored dark matter in the literature are special cases that fall into
one of these categories. We then focus on a specific class of models
where dark matter carries tau flavor, and perform a careful detector
level study of its prospects for discovery at the Large Hadron Collider
(LHC).

To incorporate three flavors of the dark matter field, the flavor
symmetry of the SM is extended from U(3$)^5$ to U(3$)^5 \times$
U(3$)_{\chi}$, if $\chi$ is a complex field such as a complex scalar,
Dirac fermion or complex vector boson. If instead $\chi$ is a real
field, such as a real scalar, Majorana fermion or real vector boson the
flavor symmetry is extended from U(3$)^5$ to U(3$)^5 \times$
O(3$)_{\chi}$. The new flavor symmetry U(3$)_{\chi}$ (or O(3$)_{\chi}$)
may be exact, or it may be explicitly broken as in the SM.

\begin{figure*}[htp]
  \subfloat[]{
  \includegraphics[scale=1.0]{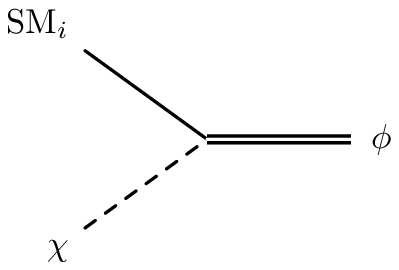}}
  \qquad\qquad\qquad\qquad\qquad
  \subfloat[]{
  \includegraphics[scale=1.0]{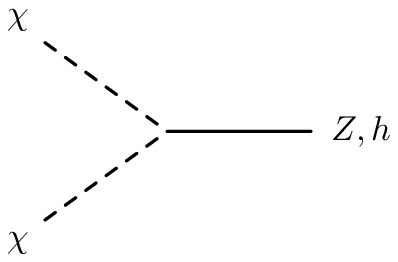}
  \qquad
  \includegraphics[scale=1.0]{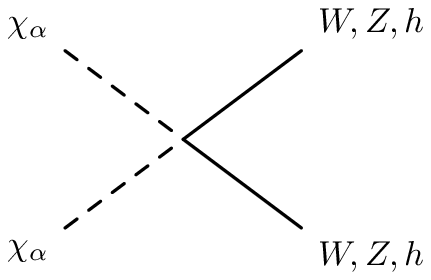}}
  \caption{Vertices that link dark matter to
  a) the SM matter fields and b) the SM gauge and Higgs fields.}
  \label{fig:dmsmint}
\end{figure*}

Our focus in this paper will be on theories where dark matter has
renormalizable contact interactions with the SM fields. Consider first
the case where these contact interactions include couplings to the SM
matter fields. These must be of the form shown in
Fig.~\ref{fig:dmsmint}(a). If the dark matter flavor symmetry is to be
exact, the field $\phi$ that mediates this interaction must transform
under U(3$)_{\chi}$ (or O(3$)_{\chi}$). If this vertex is to respect the
SM flavor symmetry, $\phi$ must also transform under the SM flavor
group. In such a scenario, the different flavor states in the dark
matter multiplet are degenerate, and the observed dark matter in the
universe will in general consist of all three flavors.

Alternatively, this contact interaction, in analogy with the SM Yukawa
couplings, could represent an explicit breaking of the flavor symmetry.
It is this scenario that we will be primarily concerned with in this
paper. In this case the simplest possibility is that the mediator $\phi$
is a singlet under both the SM and the dark matter flavor groups. Then,
if the SM matter field that $\chi$ couples to is a lepton, there is an
association between the different dark matter flavors and lepton
flavors.  Accordingly, we refer to this scenario as `lepton flavored
dark matter'. Sneutrino dark matter and Kaluza-Klein neutrino dark
matter are special cases that fall into this category, as does the model
of~\cite{Cui:2011qe}. Similarly, we label the corresponding case where
$\chi$ couples to a quark as `quark flavored dark matter'. The models of
flavored dark matter studied in~\cite{Kamenik:2011nb} fall into this
category. In this framework the fact that the SM flavor symmetries are
not exact naturally results in a splitting of the states in the dark
matter multiplet, the physical dark matter particle being identified
with the lightest.

A different class of theories involves models of flavored dark matter
where direct couplings between $\chi$ and the SM matter fields at the
renormalizable level are absent. Instead, the contact interactions of
$\chi$ with SM fields are either with the $W$ and $Z$ gauge bosons, or
with the Higgs, and can naturally preserve both the SM flavor symmetries
and the dark matter flavor symmetry. The general form of such vertices
is shown in Fig.~\ref{fig:dmsmint}(b). Closely related to this are
theories with interactions of the exactly same form, but where dark
matter instead couples to a new scalar $\phi$ or vector boson $Z'$,
which then acts as a mediator between the SM fermions and the dark
sector. The model of flavored dark matter studied in ~\cite{Kile:2011mn}
falls into this category. In such a framework, dark matter is in general
not associated with either quark or lepton flavor. We therefore refer to
this scenario as `internal flavored dark matter'.

Since the characteristic vertices of lepton flavored, quark flavored
and internal flavored dark matter are distinct, their implications for
phenomenology are very different. In the next section we consider each
of these classes of theories in turn, and study their collider
signals, as well as their implications for direct detection and flavor
physics. We then focus on a specific model of tau flavored dark matter
and show that its collider signals include events with four or more
isolated leptons and missing energy that can allow these theories to
be discovered at the LHC above SM backgrounds. We also study the
extent to which flavor and charge correlations among the final state
leptons allows models of this type to be distinguished from more
conventional theories where the dark matter particle couples to
leptons but does not carry flavor, such as neutralino dark matter in
the MSSM.

\section{Flavored Dark Matter}

\subsection{Lepton Flavored Dark Matter}

We first consider the case where dark matter carries lepton flavor. The
lepton sector of the SM has a U(3$)_{L}$ $\times$ U(3$)_{E}$ flavor
symmetry, where U(3$)_{L}$ acts on the SU(2) doublet leptons and
U(3$)_{E}$ on the singlets. This symmetry is explicitly broken down to
U(1$)^3$ by the Yukawa interactions that give the charged leptons their
masses. (We neglect the tiny neutrino masses, which also break the
symmetry). The characteristic vertex of lepton flavored dark matter
involves contact interactions between $\chi$ and the SM leptons of the
form shown in the Fig.~\ref{fig:dmsmint}(a). The corresponding terms in
the Lagrangian take the schematic form
\begin{align}
  {\lambda_{A}}^\alpha L^{A} \chi_{\alpha} \phi \; \; + {\rm h.c.},
\end{align}
if dark matter couples to the SU(2) doublet leptons $L$ of the SM
or alternatively,
\begin{align}
\label{leptonsinglet}
{\lambda_{\alpha}}^ i \chi^{\alpha} E^c_i \phi \; \; + {\rm h.c.},
\end{align}
if dark matter couples to the SU(2) singlet leptons $E^c$. Here $A$
is a U(3$)_L$ flavor index while $i$ is a U(3$)_E$ flavor index and
$\alpha$ is a U(3$)_{\chi}$ flavor index. There may also be additional
interactions between the dark matter fields and the SM of the form shown
in Fig.~\ref{fig:dmsmint}(b), in particular when $\chi$ transforms
under the SU(2) gauge interactions of the SM.

\begin{figure}[tp]
    \includegraphics[scale=1.2]{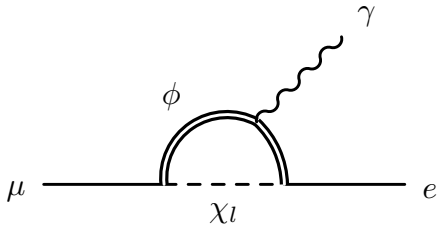}
  \caption{Potential contribution to $\mu\rightarrow e \gamma$ from
  lepton flavored dark matter.}
  \label{fig:leptonflavor}
\end{figure}

The particle $\phi$ that mediates dark matter interactions with the
charged leptons is necessarily electrically charged. If $\chi$ is a
fermion then $\phi$ must be a boson and vice versa. Any symmetry that
keeps $\chi$ stable will carry over to $\phi$, and so $\phi$ cannot
decay entirely into SM states, except perhaps on cosmological
timescales if the symmetry is not exact but very weakly broken. The
same symmetry ensures that lepton flavor violating processes involving
the dark matter field, such as $\mu \rightarrow e \gamma$, only arise
at loop level through diagrams such as the one shown in Fig.
\ref{fig:leptonflavor}. For concreteness, in what follows we take
$\chi$ to be a Dirac fermion and $\phi$ to be a complex scalar, and
restrict our focus to the case where $\chi$ couples to the SU(2)
singlet lepton field $E^c$, as in Eq. \ref{leptonsinglet}. The
generalization to the other cases is straightforward, and is left for
future work.
\subsubsection{Flavor Structure}

In general the matrix $\lambda$ will contain both diagonal and
off-diagonal elements, thereby giving rise to lepton flavor violation.
The experimental bounds on such processes are satisfied if all the
elements in $\lambda \lesssim 10^{-3}$ for $m_{\phi} \sim 200$ GeV,
even in the absence of any special flavor structure. In spite of these
small couplings such a theory can still lead to interesting collider
signals, since $\phi$ can be pair produced through SM gauge
interactions, and will emit charged leptons as it decays down to the
dark matter particle. Unfortunately, however, couplings of this size
are by themselves too small to generate the correct abundance for
$\chi$, if it is to be a thermal relic. This is not necessarily a
problem if $\chi$ transforms under the SU(2) gauge interactions of the
SM, or more generally if the theory has additional vertices of the
form shown in Fig.~\ref{fig:dmsmint}(b), since these other couplings
can play a role in determining the relic abundance. However, if $\chi$
is a SM singlet and has no sizable couplings beyond those in Eq.
\ref{leptonsinglet}, the elements in $\lambda$ must be of order unity
to generate the observed amount of dark matter, and aligned with the
lepton Yukawa couplings to avoid flavor bounds.

The matrix $\lambda$ can naturally be aligned with the SM Yukawa
couplings if this interaction preserves a larger subgroup of the SM
flavor group than just overall lepton number. For example, if we
identify the three flavors of dark matter with the electron, muon and
tau flavors in the SM, alignment is obtained if $\lambda$, and the dark
matter mass matrix, respects the U(1$)^{3}$ symmetry of lepton sector of
the SM. In other words, the U(3$)_{\chi} \times$ U(3$)^2$ symmetry is
explicitly broken by $\lambda$, and by the SM Yukawa couplings, down to
the diagonal U(1$)^{3}$. This larger symmetry forbids lepton flavor
violating processes.

A more restrictive possibility is that the only sources of flavor 
violation in the theory are the SM Yukawa couplings. This then 
constrains the matrix $\lambda$ to be consistent with minimal flavor 
violation (MFV)~\cite{D'Ambrosio:2002ex}. In this scenario, the dark 
matter flavor symmetry U(3$)_{\chi}$ is identified with either 
U(3$)_{E}$ or U(3$)_{L}$ of the SM, and the matrix $\lambda$ respects 
these symmetries up to effects arising from the SM Yukawa couplings.

If we write the lepton Yukawa couplings of the SM as
\begin{align}
{y_A}^i L^A E^c_i H \; \; + {\rm h.c.},
\end{align}
then the Yukawa matrix ${y_A}^i$ can be thought of as a spurion
transforming as $(3,\bar{3})$ under the SU(3$)_L \times $ SU(3$)_E$ subgroup
of U(3$)_L \times $ U(3$)_E$.
Consider first the case where U(3$)_{\chi}$ is
identified with U(3$)_{E}$. Then
\begin{align}
{\lambda_{\alpha}}^i \chi^{\alpha} E^c_i \phi \; \; + {\rm h.c.}
\rightarrow {\lambda_j}^i \chi^{j} E^c_i \phi \; \; + {\rm h.c.}
\end{align}
If the theory respects MFV the matrix $\lambda$ is
restricted to be of the form
\begin{align}
  {\lambda_j}^i = {\left( \alpha \mathds{1} + \beta \; y^{\dagger} y \right)_j}^i \; .
\label{lambdaexpansion}
\end{align}
Here $\alpha$ and $\beta$ are constants, and we are keeping only the
first non-trivial term in an expansion in powers of the SM Yukawa
couplings.

We write the dark matter mass term schematically as
\begin{equation}
\left[ m_{\chi} \right]{_\beta}^\alpha \bar{\chi}_{\alpha} \chi^{\beta} ,
\end{equation}
In this case MFV restricts $m_{\chi}$ to have the form
\begin{align}
\left[ m_{\chi} \right]{_i}^j
= {\left( m_0 \mathds{1} + \Delta m \; y^{\dagger} y \right)_i}^j .
\label{mexpansion}
\end{align}
where $m_0$ and $\Delta m$ are constants.

The spectrum and phenomenological implications arising from this
scenario depend sensitively on the values of the parameters $\beta$ and
$\Delta m$. In any specific model, these constants will depend on
details of the underlying ultraviolet physics. However, we expect that
in the absence of tuning, any theory where the Yukawa couplings
constitute a sufficiently small breaking of the flavor symmetry that
perturbative expansions of the matrices $\lambda$ and $m_{\chi}$ in
powers of the Yukawa couplings, as in Eqs.~\ref{lambdaexpansion}
and~\ref{mexpansion}, are justified will satisfy the inequalities
 \begin{eqnarray}
\label{inequalitiesforvalidity}
\alpha &\gtrsim& \beta y_{\tau}^2 \nonumber \\
m_0 &\gtrsim& \Delta m y_{\tau}^2  \; .
 \end{eqnarray}
Here $y_{\tau}$ is the Yukawa coupling of the tau lepton in the SM.

Since the SM Yukawa couplings of the first two generations are very 
small, we see that the corresponding dark matter flavors have very small 
splittings and couple in a flavor diagonal way with approximately equal 
strength to leptons of the SM. Depending on the sign of $\Delta m$, 
either the tau flavored or the electron flavored state will be the 
lightest. For dark matter masses of order 100 GeV, these mass splittings 
are a few hundred MeV or less. The tau flavored dark matter state can, 
however, be split from the other two states by up to tens of GeV. The 
strength of its couplings to the SM may also be somewhat different from 
the other flavors.

We now turn to the case where U(3$)_{\chi}$ is identified with U(3$)_{L}$.
Then
\begin{align}
{\lambda_{\alpha}}^ i \chi^{\alpha} E^c_i \phi \; \; + {\rm h.c.}
\rightarrow {\lambda_A}^i  \chi^{A} E^c_i \phi \; \; + {\rm h.c.}
\end{align}
MFV restricts the matrix $\lambda$ to be of the form
\begin{align}
{\lambda_A}^i = {\kappa} \; {y_A}^i \; ,
\end{align}
where ${\kappa}$ is a constant. Again we are working only to the
leading non-trivial order in an expansion in the SM Yukawa couplings.
The dark matter mass term now takes the form
 \begin{align}
\left[{m_{\chi}}\right]{_A}^B
= {\left( m_0 \mathds{1} + \Delta m \; y y^{\dagger} \right)_A}^B .
 \end{align}
While we still require $m_0 \gtrsim \Delta m y_{\tau}^2$ for consistency,
there is no corresponding constraint on ${\kappa}$ beyond requiring
that the ${\lambda_A}^i$ be small enough for perturbation theory to be
valid. We see that as in the previous case the electron and muon dark
matter flavors are necessarily close in mass, while for large values of
$\Delta m$ the tau flavor can be somewhat split. However, the couplings
of the different dark matter flavors to the SM fields, though still
flavor-diagonal, are now hierarchical. In particular, if the relic
abundance is determined by $\lambda$, and if the splitting between the
tau flavor and the other two is much larger than the temperature at
freeze out, we expect that only the tau flavor can constitute dark
matter, since the couplings of the other flavors are too small to
generate the correct abundance.

\begin{figure*}[tp]
    \begin{center}
    \includegraphics[scale=1.2]{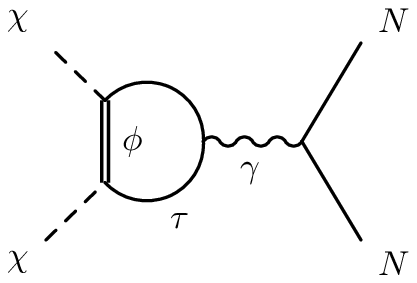}
    \qquad\qquad
    \qquad
    \includegraphics[scale=1.2]{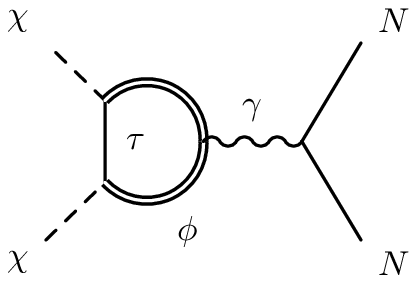}
  \end{center}
  \caption{Dark matter scattering off a nucleus through photon
  exchange.}
  \label{fig:ddleptons}
\end{figure*}

\subsubsection{Relic Abundance}

If $\chi$ is a thermal WIMP, its relic abundance is set by its 
annihilation rate to SM fields. We will concentrate on the case where 
$\chi$ is a SM singlet, and its only interactions are those of 
Eq.~\ref{leptonsinglet}. Then the primary annihilation mode is through 
$t$-channel $\phi$ exchange to two leptons. In the relevant parameter 
space, the matrix $\lambda$ is constrained by flavor bounds to be very 
nearly flavor diagonal, so each state in the dark matter multiplet is 
associated with a specific lepton flavor. We begin by assuming that the 
splittings between the different states in this multiplet are large 
enough that the heavier states do not play a significant role in 
determining the relic abundance of the lightest state. We will relax 
this assumption later.

The relevant terms in the Lagrangian, written schematically in
4-component Dirac notation, take the form
 \begin{align}
\mathcal{L}
&\supset
\frac{\lambda}{2}
\left[\bar{\chi} (1+\gamma_5)\ell \;\phi
+\bar{\ell} (1-\gamma_5) \chi\; \phi^\dagger
\right] \; .
\label{lepton4comp}
 \end{align}
Here $\chi$ represents the physical dark matter state and $l$ the
corresponding lepton. We have suppressed flavor indices since the matrix
$\lambda$ is constrained to be nearly diagonal in the relevant region of
parameter space. Since the dark matter particle is non-relativistic at
freeze-out, annihilation is dominated by the lowest partial wave. In
this limit
\begin{align}
  \langle\sigma v\rangle
  &=
  \frac{\lambda^4 m_\chi^2}{32\pi (m_\chi^2+m_\phi^2)^2},
\label{annihilationcrosssection}
 \end{align}
where we have assumed that $m_{\chi}\gg m_{\ell}$, so that the masses of
the final state leptons can be neglected.

The relic abundance is determined by solving the Boltzmann equation for
the dark matter number density $n$ at late times,
 \begin{equation}
\frac{dn}{dt} + 3 Hn = -\langle \sigma v \rangle
\left(n^2 - n_{eq}^2 \right) \;.
 \end{equation}
Here $H$ is the Hubble constant and $n_{eq}$ is the equilibrium
number density of $\chi$. In the limit that $m_{\phi} \gg m_{\chi}$,
the constraint that the relic abundance agree with observation
determines $\lambda/m_{\phi}$ as a function of the dark matter mass.

If the splitting between the different flavors of dark matter is
sufficiently small, more than one dark matter species may be present at
freeze out. In this case co-annihilations play a significant role, and
must be taken into account in the relic abundance calculation. For
concreteness, we focus on the MFV scenarios considered in the previous
section. We first consider the case where dark matter transforms under
U(3$)_E$, and where the splittings are such that the electron and muon
flavors of dark matter are both present during freeze out, but not the
tau flavor. Depending on the splittings, the muon flavored state may
either subsequently decay to the electron flavored state, or remain
stable on cosmological time scales so as to constitute a component of
the observed dark matter. In either case the number density of dark
matter is unaffected.

In this framework, the cross sections for $\chi_e \chi_{\mu}$
and $\chi_{\mu} \chi_{\mu}$ annihilations are both equal to that for
$\chi_e \chi_{e}$ annihilation, given by
Eq.~\ref{annihilationcrosssection}. If the mass splitting between the
two species is much smaller than the temperature at freeze out, $n_{eq}$
is also the same for both species. If we denote the number density of
electron flavored dark matter by $n_e$, and that of the muon flavor by
$n_{\mu}$, the Boltzmann equations take the form
\begin{eqnarray}
  \frac{dn_e}{dt} + 3 Hn_e
  &=&
  -\langle \sigma v \rangle
  \left[
  \left(n_e^2 - n_{eq}^2 \right)
  + \left(n_e n_\mu - n_{eq}^2 \right)
  \right]
  \nonumber \\ && \qquad\qquad\qquad \qquad
  - \left[
  \chi_e {\rm \rightarrow} \chi_\mu \right] \nonumber
  \\
  \frac{dn_{\mu}}{dt} + 3 Hn_{\mu}
  &=&
  -\langle \sigma v \rangle
  \left[
  \left(n_{\mu}^2 - n_{eq}^2 \right)
  + \left(n_e n_\mu - n_{eq}^2 \right)
  \right]
  \nonumber \\ && \qquad\qquad\qquad \qquad
  - \left[ \chi_{\mu} {\rm \rightarrow} \chi_e \right]
  \; .
\end{eqnarray}
Here $\chi_e \rightarrow \chi_\mu$ denotes the net effect of scattering 
processes such as $ e \chi_e \leftrightarrow \mu \chi_{\mu}$ as well as 
decays and inverse decays which convert the electron flavor of dark 
matter into the muon flavor and vice versa, but leave the overall dark 
matter density $N = n_e + n_{\mu}$ unaffected. Recognizing that the 
relic abundance is set by the value of $N$ at late times, we can combine 
the equations above to obtain a single equation for $N$,
 \begin{equation}
\frac{dN}{dt} + 3 HN = -\langle \sigma v \rangle
\left(N^2 - 4 n_{eq}^2 \right) \;.
 \end{equation}

This equation has a very similar form to the Boltzmann equation for a
single dark matter species, and can be solved in exactly the same way.
We find that the relic abundance is in fact relatively insensitive to
the change in the number of dark matter species, changing by only about
5\% when other parameters are kept fixed. A very similar analysis shows
that the same conclusion holds true when the splittings are small enough
that the tau flavor of dark matter is also in the bath at freeze out. It
follows that the results from the single flavor case,
Eq.~{\ref{annihilationcrosssection}, also apply to the cases of more
than one dark matter flavor, up to fairly small corrections.

We now move on to the case where dark matter transforms under U(3$)_L$,
and all three flavors are present at freeze out. For simplicity we work
in the limit where the splitting between all the different flavors is much
smaller than the temperature at freeze out, and can be neglected. Since
the couplings of the different dark matter flavors to the corresponding
SM fermions are now hierarchical, the cross section for $\chi_{\tau}
\chi_{\tau}$ annihilation is larger by more than two orders of magnitude
than that for $\chi_{\tau} \chi_{\mu}$ annihilation, which can be
neglected. The cross sections for $\chi_{\mu} \chi_{\mu}$, $\chi_{\tau}
\chi_{e}$, $\chi_{\mu} \chi_{e}$ and $\chi_{e} \chi_{e}$ are even
smaller, and these processes can also be neglected. The relevant
Boltzmann equations then take the form
\begin{eqnarray}
  \frac{dn_e}{dt} + 3 Hn_e
  &=&  - \left[ \chi_e {\rm \rightarrow} \chi_\mu, \chi_\tau \right]
  \nonumber
  \\
  \frac{dn_{\mu}}{dt} + 3 Hn_{\mu}
  &=&
  - \left[ \chi_{\mu} {\rm \rightarrow} \chi_e, \chi_\tau \right]
  \nonumber \\
  \frac{dn_{\tau}}{dt} + 3 Hn_{\tau}
  &=&  -\langle \sigma v \rangle
  \left[\left(n_{\tau}^2 - n_{eq}^2 \right) \right]
  - \left[ \chi_{\tau} {\rm \rightarrow} \chi_{\mu ,e} \right]
  \, .
\label{coupled1}
\end{eqnarray}
The relic abundance of dark matter depends on whether processes which 
change the flavor of dark matter but conserve total dark matter number, 
such as $e \chi_e \leftrightarrow \tau \chi_\tau$, $\mu \chi_\mu 
\leftrightarrow \tau \chi_\tau$ etc. remain in equilibrium during the 
freeze out process. If this is the case the relative fractions 
$n_e/n_{\tau}$ and $n_{\mu}/n_{\tau}$ closely track their equilibrium 
value $\sim 1$. Then we can add these equations to obtain a single 
equation for the total dark matter number $N = n_e + n_{\mu} + 
n_{\tau}$.
 \begin{equation}
\frac{dN}{dt} + 3 HN = -\langle \sigma v \rangle
\left(\frac{N^2}{9} -  n_{eq}^2 \right) \;.
 \end{equation}
This equation has a very similar form to that of the Boltzmann equation 
for a single species, but for the factor of 9, which is the square of 
the number of dark matter flavors. To generate the observed dark matter 
density then requires the annihilation cross section to be a factor of 9 
larger than in the single flavor case. This implies that the correct 
relic abundance is obtained if $\lambda$ for the $\tau$ flavor is a 
factor of $\sqrt{3}$ larger than in the single flavor case, 
Eq.~{\ref{annihilationcrosssection}}. If, however, the processes which 
change dark matter flavor $e \chi_e \leftrightarrow \tau \chi_\tau$ and 
$\mu \chi_\mu \leftrightarrow \tau \chi_\tau$ go out of equilibrium much 
before $\chi_{\tau}$ freezes out, the surviving $\chi_{\mu}$ and 
$\chi_{e}$ will contribute too much to the dark matter density to be 
consistent with observations.

In general, for realistic values of the parameters, the process $\mu 
\chi_\mu \leftrightarrow \tau \chi_\tau$ will be in equilibrium during 
freeze out. However, the rate for $e \chi_e \leftrightarrow \tau 
\chi_\tau$, which, though enhanced by a Boltzmann factor, is suppressed 
by the ratio $m_e^2/m_{\tau}^2$ relative to the annihilation process 
$\chi_{\tau} \chi_{\tau} \rightarrow \tau \tau$, is generally of order 
the expansion rate at freeze out. Therefore the approximation 
$n_e/n_{\tau} = 1$ may not be valid, and cannot be used to simplify the 
coupled equations~(\ref{coupled1}). A preliminary numerical study 
nevertheless suggests that if $\lambda$ for the $\tau$ flavor is larger 
than in the single flavor case by a factor close to $\sqrt{3}$, the 
correct abundance of dark matter is indeed obtained. However, we leave a 
detailed analysis of this scenario for future work.

If $\chi$ is also charged under the SM SU(2) gauge interactions then new
annihilation channels open up. Dark matter can annihilate into two
$W$'s, two $Z$'s, and also into SM fermions through through $s$-channel
$Z$ exchange. We leave a study of this for future work.

\subsubsection{Direct Detection}

Direct detection experiments seek to observe effects arising from the
collisions of dark matter particles with ordinary matter.  Although in
theories of electron flavored dark matter, $\chi$ can scatter off
electrons at tree level, the energy transfer is generally not enough to
generate a signal in these experiments~\cite{Kopp:2009et}. Therefore we
focus on nuclear recoils.

The direct detection signals of this class of theories depend on whether
the dark matter particle $\chi$ transforms non-trivially under the SM
SU(2) gauge symmetry, or remains a SM singlet. If $\chi$ is a SM
singlet, the leading contribution to dark matter scattering off a
nucleus arises from the loop diagrams involving leptons shown in Fig.
\ref{fig:ddleptons}.

In the region of parameter space of interest to current direct detection
experiments, bounds on lepton flavor violating processes constrain the
coupling matrix $\lambda$ to be flavor diagonal. Therefore the dark
matter candidate carries the flavor of the lepton it couples to. We
first consider the case where the splittings between the different dark
matter flavors is large, so that a single flavor constitutes all the
observed dark matter. We will relax this assumption later. In this limit
the relevant terms in the Lagrangian are again those shown in
Eq.~\ref{lepton4comp}. As explained in Appendix \ref{appendix}, this
coupling gives rise to three distinct types of interactions between dark
matter and the nucleus, specifically a charge-charge coupling, a
dipole-charge coupling, and a dipole-dipole coupling.

\begin{figure*}[tp]
    \subfloat[]{
    \psfrag{K}[b]{$^{\lambda}/_{m_\phi}$ (GeV$^{-1}$)}
    \psfrag{M}[]{$m_\chi$ (GeV)}
    \includegraphics[scale=0.7]{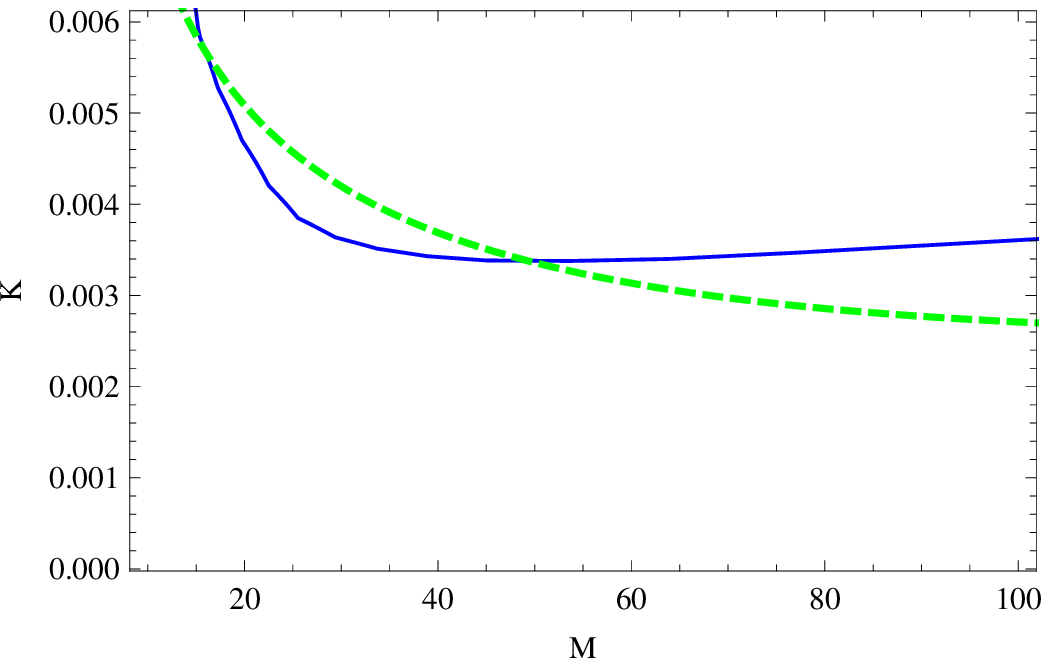}
    }
    \hspace{0.3in}
    \subfloat[]{
    \psfrag{K}[b]{$^{\lambda}/_{m_\phi}$ (GeV$^{-1}$)}
    \psfrag{M}[]{$m_\chi$ (GeV)}
    \includegraphics[scale=0.7]{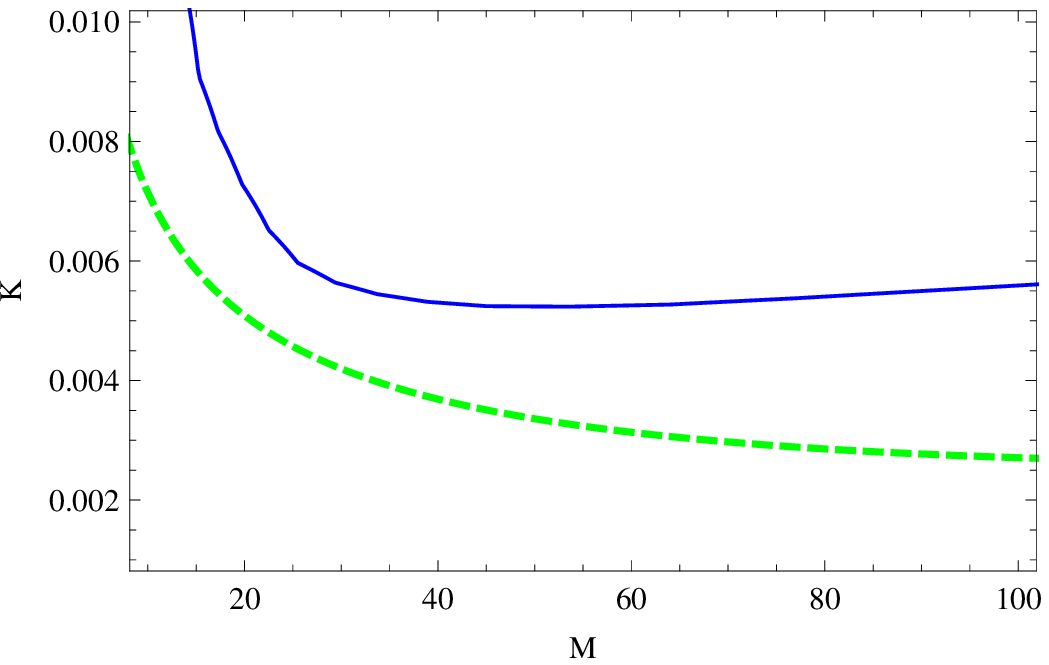}
    }
  \caption{Direct detection and
  relic abundance constraints on lepton flavor dark matter for
  a)$\chi_e$ and b)$\chi_\tau$, when $m_\phi=150$ GeV. The area above the solid
  blue curve is ruled out by the new Xenon100 \cite{Aprile:2011hi}
  data. The green dashed
  curves signify the parameters for which we obtain correct relic
  abundance. }
  \label{fig:lepton-dd-relic}
\end{figure*}

The differential cross section for the charge-charge cross section is
given by the expression,
\begin{align}
  \frac{d\sigma_{ZZ}}{dE_r}
  &=
  \frac{2 m_N}{4\pi  v^2}
  Z^2\, b_p^2\,
  F^2(E_r)
\end{align}
where $m_N$ is the mass of the nucleus, $v$ is the velocity of the
dark matter particle and $E_r$ is the recoil energy of the nucleus.
Note that this is a spin-independent
interaction, and hence is enhanced by $Z$, the total charge of the nucleus.
The form factor $F(E_r)$ appearing here is the
charge form factor of the nucleus. It has been measured explicitly
to be in good agreement with the Helm form factor \cite{Duda:2006uk}.
The coefficient $b_p$ is defined as
\begin{align}
  b_p&=
  \frac{\lambda^2 e^2}{64\pi^2 m_\phi^2}
  \left[1+\frac23\log\left(\tfrac{m_{\ell}^2}{m_\phi^2}\right) \right].
\label{eqnbp2}
\end{align}
Here $m_{\ell}$ is the mass of the lepton in the loop, which has the
same flavor as the dark matter particle. The leading logarithmic part of
this expression was calculated in \cite{Kopp:2009et}. In the case of
electron flavored dark matter, the mass of the lepton $m_{\ell}$ in
Eq.~\ref{eqnbp2} must be replaced by the momentum transfer $|\vec{k}|$
in the process, which we take to be 10 MeV as a reference value.

The magnetic dipole moment of the dark matter can also couple to the
electric charge of the nucleus. This interaction is also
spin-independent.
\begin{align}
  \frac{d\sigma_{DZ}}{dE_r}
  &=
  \frac{e^2 Z^2 \mu_\chi^2}{4\pi E_r}
  \left[
  1-\frac{E_r}{v^2}
  \frac{m_\chi+2m_N}{2m_N m_\chi}
  \right]
  F^2(E_r)
\end{align}

Finally, the dark matter can couple to the nuclear magnetic dipole moment via a
dipole-dipole coupling. This interaction is spin-dependent, and therefore
does not get an enhancement for large nuclei. It takes the form
\begin{align}
  \frac{d\sigma_{DD}}{dE_r}
  &=
  \frac{m_N\, \mu_{nuc}^2 \,\mu_\chi^2}{\pi v^2}
  \left( \frac{S_{nuc}+1}{3S_{nuc}} \right)
  F_D^2(E_r)
  .
\end{align}
where  $S_{nuc}$ is the spin of the nucleus,
$\mu_{nuc}$ is its magnetic
dipole moment, and $F_D(E_r)$ is the dipole moment form
factor for the  nucleus. There are currently no explicit measurements
of the magnetic dipole form factor. A discussion of various form
factors and an approximate calculation can be found in
\cite{Chang:2010en} (and references therein). The magnetic dipole
moment of the dark matter particle $\mu_\chi$ is related to the model
parameters by
\begin{align}
  \mu_\chi
  &=
  \frac{\lambda^2 e\, m_\chi}{64 \pi^2 m_\phi^2} \; .
\end{align}
Note that there is also a potential charge-dipole contribution to the
cross section, where the dark matter vector bilinear couples to the
magnetic dipole moment of the nucleus, but this interaction is
suppressed by additional powers of momentum transfer.

The dipole-charge interaction is sub-dominant to the charge-charge
interaction.  The dipole-dipole coupling, being spin-dependent, is
also sub-dominant. Consequently, we use the charge-charge cross
section for placing limits. Then,
 \begin{align}
  \sigma^0_{ZZ}
  &=
  \frac{\mu^2 Z^2}{\pi}
  \left[
  \frac{\lambda^2 e^2}{64\pi^2 m_\phi^2}
  \left[1+\frac23\log\left(\tfrac{m_{\ell}^2}{m_\phi^2}\right)
  \right]
  \right]^2.
\end{align}
Here $\sigma_0$ is the cross section at zero-momentum transfer, and
$\mu$ is the reduced mass of the dark matter-nucleus system.

\begin{figure*}[tp]
    \subfloat[]{
    \includegraphics[scale=1.0]{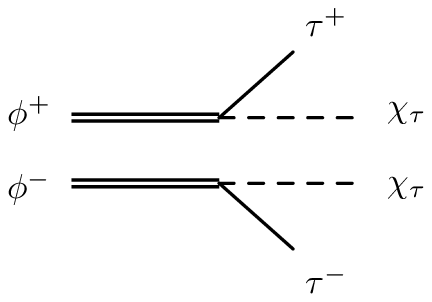}}
    \hspace{0.3in}
    \subfloat[]{
    \includegraphics[scale=1.0]{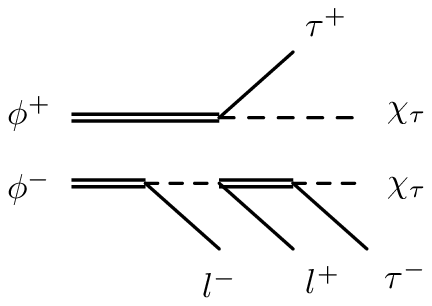}
    }
    \hspace{0.3in}
    \subfloat[]{
    \includegraphics[scale=1.0]{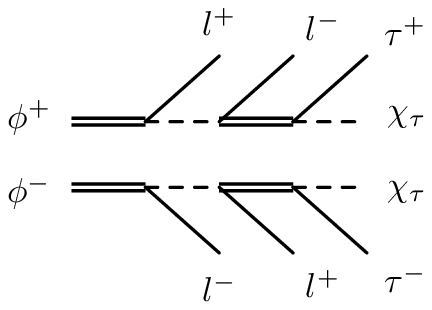}
    }
  \caption{Signal topologies at colliders in models of
  tau flavored dark matter.}
  \label{fig:topology}
\end{figure*}
The ratio $\lambda/m_{\phi}$ corresponding to a thermal WIMP is plotted
in Fig.~\ref{fig:lepton-dd-relic} as a function of the dark matter mass,
for the tau flavored and electron flavored cases. The current limits
from the Xenon100 experiment~\cite{Aprile:2011hi} are also shown. It is
clear from the figure that the expected improvement in sensitivity of
the experiment by an order of magnitude will bring a large part of the
parameter space of these models within reach.

In scenarios motivated by MFV, the splitting between the different
states in the dark matter multiplet may be small enough that more than
one state is present in the bath at freeze out. The observed dark matter
may also be composed of more than one flavor, if the splittings are
small enough that the lifetimes of the heavier flavours are longer than
the age of the universe. If all the dark matter flavors couple to the
corresponding SM particles with the same strength, as when $\chi$
transforms under U(3$)_E$, the calculation of the previous section shows
that the parameters that give rise to the observed relic abundance are
fairly insensitive to the number of dark matter species at freeze out.
If the lightest flavor, whether $\chi_e$ or $\chi_{\tau}$, is split from
the others by a few tens of MeV or more, the heavier states will decay
down to it, and the observed dark matter is composed of a single flavor.
In this scenario the direct detection bounds on $\lambda$ are
essentially unchanged. For smaller splittings, more than one flavor
could constitute the observed dark matter today. The bound in this case
may be obtained by appropriately interpolating between the somewhat
different limits in the single flavor cases.

If, however, the quasi-degenerate dark matter flavors couple 
hierarchically, as when $\chi$ transforms under U(3$)_L$, the relic 
abundance calculation of the previous section shows that $\lambda$ for 
the tau flavor is larger by a factor of $\sqrt{3}$ or more than in the 
single flavor case. For the purpose of setting a conservative bound on 
$\lambda$ for the tau flavor from direct detection, we will take this 
factor to be $\sqrt{3}$. Then, in this scenario, if the observed dark 
matter is composed of the tau flavor, the limits are stronger by 
$\sqrt{3}$ than in the single flavor case. On the other hand, if dark 
matter today is composed of the $e$ or $\mu$ flavors, the limits are 
much weaker than in the corresponding single flavor cases because of the 
hierarchical couplings of $\chi$. The lightest flavor, whether $\chi_e$ 
or $\chi_{\tau}$, will constitute all of dark matter if it is split from 
the other states by a few hundred MeV or more. For smaller splittings, 
more than one flavor may constitute the observed dark matter. The bound 
then depends on the constituent fraction of $\chi_{\tau}$, and may be 
obtained by interpolation.

If $\chi$ does transform under SU(2), we expect that the leading
contribution to the cross section for dark matter scattering off a
nucleus will arise from tree-level exchange of the SM $Z$, provided
$\chi$ carries non-zero hypercharge. If $\chi$ arises from a
representation which does not transform under hypercharge, then it does
not couple directly to the $Z$, and so this effect does not arise. In
this scenario, loop diagrams involving $W$ bosons generate a
contribution to the cross section~\cite{Cirelli:2005uq} that must be
compared against the contribution from the lepton loop above in order to
determine the leading effect.

\subsubsection{Collider Signals}

What are the characteristic collider signals associated with this class
of theories? For concreteness, we limit ourselves to the case where
$\chi$ does not transform under the SM SU(2) gauge interactions, and is
a SM singlet. Then the mediator $\phi$ also does not transform under the
SU(2) gauge symmetry.

We focus on the scenario where dark matter couples flavor diagonally,
and where the electron and muon flavored states in the dark matter
multiplet are highly degenerate, as would be expected from MFV. In such
a framework, the charged leptons that result from the decay of a muon
flavored state to an electron flavored one (or vice versa) are extremely
soft, and would be challenging to detect in an LHC environment. For the
purposes of the following discussion, we will assume that these leptons
are not detected. However, the splitting between a tau flavored state
and an electron or muon flavored one is assumed to be large enough that
the corresponding leptons can indeed be detected.

The mediators $\phi$ can be pair-produced in colliders through an
off-shell photon or $Z$. Each $\phi$ can then either decay directly to
the dark matter particle, or decay to one of the heavier particles in
the dark matter multiplet which then cascades down to the dark matter
particle. Then, if the dark matter particle carries tau flavor, the
decay of each $\phi$ results in either a single tau, or in two charged
electrons or muons and a tau. Each event is therefore associated with
exactly two taus, up to four additional charged leptons, and missing
energy. These event topologies are shown in Fig. \ref{fig:topology}.
If, on the other hand, the dark matter particle carries electron
flavor, the decay of each $\phi$ will result in either a solitary electron
or muon, or in two taus and an electron or muon. We therefore expect two
electrons, two muons or an electron and a muon in each event, along with
missing energy and up to four taus.

\subsection{Quark Flavored Dark Matter}

Let us now consider the case where dark matter carries quantum numbers under
quark flavor. The quark sector of the SM has a U(3$)_{Q}$ $\times$
U(3$)_{U}$ $\times$ U(3$)_{D}$ flavor symmetry, where U(3$)_{Q}$ acts on
the SU(2) doublet quarks and U(3$)_{U}$ and U(3$)_{D}$ on the up and
down-type singlet quarks. This symmetry is explicitly broken down to
U(1) baryon number by the SM Yukawa couplings.

The characteristic vertex of quark flavored dark matter has the form
shown in Fig.~\ref{fig:dmsmint}(a). The corresponding terms in the
Lagrangian take the schematic form
\begin{align}
{\lambda_{A}}^{\alpha} Q^A \chi_{\alpha}\; \phi \; \; + {\rm h.c.} \; ,
\label{quarkdoublet}
\end{align}
if dark matter couples to the SU(2) doublet quarks $Q$. Alternatively,
if it couples to the SU(2) singlet up-type quarks $U^c$, we have
\begin{align}
{\lambda_{\alpha}}^{i} \chi^{\alpha} U^c_i \; \phi \; \; + {\rm h.c.}
\label{quarksingletup}
\end{align}
This is easily generalized to the case where dark matter transforms
under U(3$)_D$.
\begin{align}
{\lambda_{\alpha}}^{a} \chi^{\alpha} D^c_a \; \phi \; \; + {\rm h.c.}
\label{quarksingletdown}
\end{align}
Here the index $A$ represents a U(3$)_Q$ flavor index while $i$ is a
U(3$)_U$ flavor index and $a$ is a U(3$)_{D}$ flavor index. The mediator
$\phi$ is now charged under both color and electromagnetism. For
concreteness, in what follows we again take $\chi$ to be a Dirac fermion
and $\phi$ to be a complex scalar, and restrict our focus to the cases
where $\chi$ couples to the SU(2) singlet quarks $U^c$ or $D^c$ as in
Eq.~\ref{quarksingletup} and Eq.~\ref{quarksingletdown}. The
generalization to other cases is straightforward, and is left for future
work.

\subsubsection{Flavor Structure}
Contributions to flavor violating processes, such as $K - \bar{K}$
mixing, arise at loop level through diagrams such as the one in Fig.
\ref{fig:quarkflavor}. The experimental bounds on flavor violation are
satisfied if all the elements in $\lambda \lesssim 10^{-2}$, for
$m_{\phi} \sim 500$ GeV. However, as in the lepton case, couplings of
this size are by themselves too small to generate the correct abundance
for $\chi$, if it is to be a thermal relic. This is not necessarily a
problem if $\chi$ has additional interactions with the SM, since these
may set the relic abundance. However, if $\chi$ is a SM singlet and has
no other sizable couplings, the elements in $\lambda$ must be of order
unity to generate the observed amount of dark matter. In this case the
interaction matrix $\lambda$ must be aligned with the SM Yukawa
couplings if the flavor constraints are to be satisfied.

For the matrix $\lambda$ to be naturally aligned with the SM Yukawa
couplings this interaction must preserve, at least approximately, a
larger subgroup of the SM flavor group than just baryon number. This
constraint is satisfied if the couplings $\lambda$ are consistent with
MFV. In this framework, the only sources of flavor violation are the SM
Yukawa couplings, and the matrix $\lambda$ respects the SM flavor
symmetries up to effects that arise from them.

Consider first the case when dark matter couples to the up-type
SU(2) singlet quarks $U^c$ as in Eq.~\ref{quarksingletup}. MFV can be
realized if the dark matter flavor symmetry U(3$)_{\chi}$ is identified
with one of U(3$)_{U}$, U(3$)_{Q}$ or U(3$)_{D}$ of the SM, and the
matrix $\lambda$ respects these symmetries up to effects arising from
the SM Yukawa couplings. As in the lepton case, we will work to the
leading non-trivial order in an expansion in powers of the SM Yukawa
couplings.

\begin{figure}[tp]
    \includegraphics[scale=1.2]{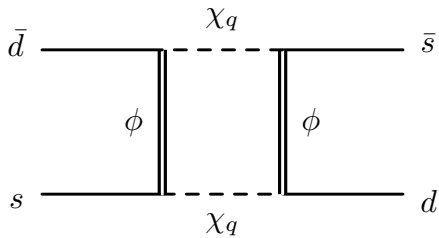}
    \caption{Potential contribution to $K-\bar{K}$ mixing from quark
    flavored dark matter.}
  \label{fig:quarkflavor}
\end{figure}
The quark Yukawa couplings in the SM can be written as
\begin{align}
  {\hat{y}_A}^{\ \ a} Q^A D^c_a H + {y_A}^i Q^A U^c_i H^{\dagger}
\; \; + {\rm h.c.} \; .
\end{align}
The up-type Yukawa matrix $y$ can be thought of as a spurion
transforming as $(3,\bar{3},1)$ under the SU(3$)_Q \times $ SU(3$)_U
\times $ SU(3$)_D$ subgroup of U(3$)_Q \times $ U(3$)_U \times $
U(3$)_D$, while the down-type matrix $\hat{y}$ can be thought of as a
spurion transforming as $(3,1,\bar{3})$.

If U(3$)_{\chi}$ is identified with U(3$)_{U}$, MFV restricts the matrix
$\lambda$ to be of the form
 \begin{align}
   {\lambda_i}^{j} = {\left( \alpha \mathds{1} + \beta \; y^{\dagger} y \right)_i}^j \; .
\end{align}
while the mass term for $\chi$ becomes
\begin{align}
\left[m_{\chi}\right]_i^j =
{\left( m_0 \mathds{1} + \Delta m \; y^{\dagger} y \right)_i}^j .
 \end{align}
As explained earlier, we are working to leading non-trivial order in
an expansion in powers of the SM Yukawa couplings. We expect that in any
theory where the Yukawa couplings constitute a sufficiently small
breaking of the flavor symmetry that such expansions of $\lambda$ and
$m_{\chi}$ in powers of the Yukawa couplings are permitted, in the absence
of tuning the inequalities
 \begin{eqnarray}
\alpha &\gtrsim& \beta y_t^2 \nonumber \\
m_0 &\gtrsim& \Delta m y_t^2 \; ,
\label{qrestriction}
 \end{eqnarray}
will be satisfied. Here $y_t$ is the Yukawa coupling of the top in the
SM. The effect of these inequalities is to constrain the mass splittings
between the different dark matter flavors, and to restrict the extent to
which their couplings can differ.

It follows from this discussion that the dark matter states that couple
to the first two generations of SM quarks are nearly degenerate in mass,
and the mixing between them and the state with top flavor is small,
protecting against flavor violating processes. For a dark matter mass of
100 GeV, the splitting between the up and charm flavored dark matter
states is less than 10 MeV. The splitting between these states and the
top flavored state can however be significantly larger, as much as tens
of GeV. The physical dark matter particle is expected to be either up
flavored or top flavored, depending on the sign of $\Delta m$.

If U(3$)_{\chi}$ is identified with U(3$)_Q$ we have instead
\begin{align}
{\lambda_A}^i = \kappa \; {y_A}^i .
\end{align}
The dark matter mass term now takes the form
\begin{align}
\left[{m_{\chi}}\right]{_A}^B
= {\left( m_0 \mathds{1} + \Delta m \; y y^{\dagger} +
\hat{\Delta} m \; \hat{y} \hat{y}^{\dagger} \right)_A}^B .
 \end{align}
The consistency of our expansion in powers of the Yukawa couplings
requires that the inequalities $m_0 \gtrsim \hat{\Delta} m y_b^2$ and
$m_0 \gtrsim {\Delta} m y_t^2$ be satisfied. Here $y_b$ is the bottom
Yukawa coupling in the SM. While the first two flavors of $\chi$ are
again quasi-degenerate in mass, their couplings to the SM are now
hierarchical rather than universal. For a dark matter mass of 100 GeV,
the dark matter flavors associated with the first two generations are
split by less than or of order 100 MeV. The third generation dark matter
particle could, however, be split from the others by tens of GeV. It is
the smallness of the SM Yukawa couplings of the first two generations
and their small mixing with the third generation that protects against
flavor changing processes. If the splittings between the third
generation dark matter particle and the other two flavors is much larger
than the temperature at freeze out, we expect that the observed dark
matter will belong to the third generation, since the other flavors
couple too weakly to give rise to the observed relic abundance.

Finally, if U(3$)_{\chi}$ is identified with U(3$)_D$ we have
\begin{align}
{\lambda_a}^i = \hat{\kappa} \; {\left(\hat{y}^{\dagger} y \right)_a}^i
\end{align}
and
 \begin{align}
{\left[m_{\chi}\right]_a}^b = {\left( m_0 \mathds{1} +
\hat{\Delta} m \;  \hat{y}^{\dagger} \hat{y} \right)_a}^b \; .
 \end{align}
For consistency we require $ m_0 \gtrsim \hat{\Delta} m y_b^2 \; $. Once
again the first two flavors are very close in mass, and their couplings
to the SM hierarchical. For a dark matter mass of order 100 GeV their
splitting is expected to be less than or of order 100 MeV. The mass of
the bottom flavored dark matter state can be split from the other two
flavors by tens of GeV. If the splittings between the bottom flavored
state and the others are much larger than the temperature at freeze out,
the lightest particle must be bottom flavored to generate the observed
abundance of dark matter.

We now turn our attention to the case where dark matter couples to the
SU(2) singlet down-type quarks $D^c$ as in Eq.~\ref{quarksingletdown}. MFV
can be realized if the dark matter flavor symmetry U(3$)_{\chi}$ is
identified with one of U(3$)_{D}$, U(3$)_{Q}$ or U(3$)_{U}$ of the SM.
The corresponding formulas for the form of the coupling matrix $\lambda$
and the dark matter mass may be obtained by simply interchanging $y$ and
$\hat{y}$ in the equations above. If U(3$)_{\chi}$ is identified with
U(3$)_{D}$, the matrix $\lambda$ is constrained to be of the form
  \begin{align}
{\lambda_a}^{b} =
{\left( {\alpha} \mathds{1} +
{\beta} \; \hat{y}^{\dagger} \hat{y} \right)_a}^b \; .
 \end{align}
while the dark matter mass term is now
 \begin{align}
{\left[m_{\chi}\right]_a}^b =
{\left( m_0 \mathds{1} + \hat{\Delta} m \;
\hat{y}^{\dagger} \hat{y} \right)_a}^b .
 \end{align}
Rather than Eq.~\ref{qrestriction}, we now have
 \begin{eqnarray}
\alpha &\gtrsim& \beta y_b^2 \nonumber \\
m_0 &\gtrsim& \hat{\Delta} m y_b^2 \; .
 \end{eqnarray}
We see that the states associated with the first two generations are
quasi-degenerate in mass and couple universally to the SM, while the
third generation can be somewhat split. This fact, together with the
small mixing between the third flavor of dark matter and the first two
allows flavor constraints to be satisfied. We expect that either the
bottom or down flavor will constitute dark matter.

If U(3$)_{\chi}$ is instead identified with U(3$)_{Q}$, we have
 \begin{align}
  {\lambda_A}^a = \kappa \; {\hat{y}_A}^{\ \ a} \; ,
 \end{align}
while the mass term is of the form
 \begin{align}
\left[{m_{\chi}}\right]{_A}^B
= {\left( m_0 \mathds{1} + \Delta m \; y y^{\dagger} +
\hat{\Delta} m \; \hat{y} \hat{y}^{\dagger} \right)_A}^B \; .
 \end{align}
We expect that the parameters will satisfy the inequalities $ m_0
\gtrsim \Delta m y_t^2 $ and $m_0 \gtrsim \hat{\Delta} m y_b^2 $. While
the first two generations are still nearly degenerate, the different flavors
now couple to the SM hierarchically rather than universally. As a
consequence we expect that if dark matter is a thermal relic, and the
splitting between the different flavors of $\chi$ much larger than the
temperature at freeze out, the observed dark matter will be composed of
third generation particles.

Finally, if U(3$)_{\chi}$ is identified with U(3$)_{U}$ these formulae
become
\begin{align}
{\lambda_i}^a = \hat{\kappa} \; {\left({y}^{\dagger} \hat{y} \right)_i}^a
\end{align}
and
\begin{align}
{\left[m_{\chi}\right]_i}^j =
{\left( m_0 \mathds{1} + \Delta m \; y^{\dagger} y \right)_i}^j \; .
\end{align}
The parameters must satisfy the inequality $ m_0 \gtrsim \Delta m y_t^2
$. Once again the dark matter states associated with the first two
generations are very close in mass, and their couplings to the SM
hierarchical. If the splitting between the top flavored state and the
other two states is much larger than the temperature at freeze out, the
observed dark matter must be top flavored,

\subsubsection{Relic Abundance}

If the primary interaction of dark matter with the SM is through
Eq.~\ref{quarksingletup} or Eq.~\ref{quarksingletdown}, then the relic
abundance is set by $t$-channel annihilation to quarks. The calculation
in this case mirrors that of the lepton flavored dark matter. In cases
where the dominant annihilation mode is kinematically forbidden (e.g.
for the top quark), three-body final states or loop-suppressed processes
may dominate.

In the region of parameter space which gives rise to the observed relic
abundance, constraints on flavor changing neutral current processes
require that the interaction matrix $\lambda$ be closely aligned with
the quark Yukawa couplings. We therefore limit our analysis to the case
where $\lambda$ is consistent with MFV. Then, in the mass basis each
particle in the dark matter multiplet is associated with the flavor of
the quark it couples to most strongly, and does not mix significantly
with the other flavors. We begin by considering the case when the
splittings between the particles in the dark matter multiplet are large
enough that only the lightest state is present in the bath on the time
scales when freeze out occurs. We will relax this assumption later. The
relevant terms in the Lagrangian, written in Dirac 4-component notation,
take the schematic form
 \begin{align}
\mathcal{L}
&\supset
\frac{\lambda}{2}
\left[\bar{\chi} (1+\gamma_5)q \;\phi
+\bar{q} (1-\gamma_5) \chi\; \phi^\dagger
\right] \; .
\label{quarkdirect}
 \end{align}
Here $\chi$ represents the physical dark matter particle and $q$ the
corresponding quark. MFV ensures that the coupling matrix $\lambda$ is
flavor diagonal in the quark mass basis, allowing us to suppress flavor
indices.  In the limit that the masses of the final state quarks can be
neglected, we find for the annihilation rate
 \begin{align}
  \langle\sigma v\rangle
  &=
  \frac{3\lambda^4 m_\chi^2}{32\pi (m_\chi^2+m_\phi^2)^2} \; .
 \end{align}
The additional factor of $3$ relative to the lepton case arises because of the
three colors of quarks. The relic abundance can then be determined from the
Boltzmann equation, and $\lambda/m_{\phi}$ determined as a function of the
dark matter mass.

The splitting between the different dark matter flavors may be small
enough that more than one species is present in the bath during freeze
out. In particular, it follows from our MFV analysis that if the
lightest dark matter particle carries either the up or down flavor, the
splitting between it and the nearest state is expected to be small
enough that both species are present in the bath during freeze out. For
some range of parameters, the splittings are such that all three flavors
are present. In such a scenario, coannihilations are expected to play a
significant role, and will affect the relic abundance of dark matter.

For concreteness, we focus on the realizations of MFV where dark matter
couples to the down-type SU(2) singlet quarks $D^c$ as in
Eq.~\ref{quarksingletdown}. We first consider the case where $\chi$
transforms under U(3$)_D$ and the lightest state carries down flavor. The
splittings are assumed to be such that both the down and strange flavored
states are in the bath at freeze out. In this realization of MFV, the
different flavors of dark matter couple with equal strength to the
associated SM particles, and so the cross sections for $\chi_d \chi_s$
and $\chi_s \chi_s$ annihilation are equal to that for $\chi_d \chi_d$
annihilation. Then an analysis very similar to that in the lepton case
shows that if all other parameters are kept fixed, the presence of the
extra quasi-degenerate species only alters the relic abundance by about
5\%. Therefore the parameters that generate the correct relic abundance
in the case of quasi-degenerate dark matter flavors differ only slightly
from the corresponding parameters in the non-degenerate case. The same
conclusion holds if the bottom flavored state is also in the bath at freeze
out.

We move on to the scenario where $\chi$ transforms under U(3$)_Q$. In 
this case the couplings of $\chi$ are hierarchical, and so if the 
lightest state is associated with the first generation, it must be 
quasi-degenerate with the others to obtain the correct relic abundance. 
More precisely, the splittings between the different flavors must be 
small or comparable to the freeze out temperature. For concreteness we 
focus on the case where the splittings are negligible compared to the 
freeze out temperature. Then for realistic values of the parameters the 
processes $d \chi_d \leftrightarrow b \chi_b $ and $s \chi_s 
\leftrightarrow b \chi_b $, which convert one flavor of dark matter into 
another, are in equilibrium at freeze out. Then an analysis identical to 
that in the leptonic case shows that the correct relic abundance is 
obtained if the coupling $\lambda$ for the third generation is larger by 
a factor of $\sqrt{3}$ than in the single flavor case. 

If dark matter transforms under the SM SU(2) gauge interactions, other
annihilation modes open up, and may play the dominant role in
determining the relic abundance. We leave this possibility for
future work.

\subsubsection{Direct Detection}

The direct detection signals of this class of models depend on the
flavor of quark that the dark matter particle couples to, and on whether
or not $\chi$ transforms under the SM SU(2) gauge symmetry. Consider
first the case where $\chi$ is a SM singlet. In the region of parameter
space relevant to current direct detection experiments, the matrix
$\lambda$ is constrained by flavor bounds to be closely aligned with the
SM Yukawa couplings. We therefore concentrate on the case where
$\lambda$ is consistent with MFV. The relevant terms in the Lagrangian
are then again those in Eq.~\ref{quarkdirect}.

\begin{figure}[tp]
  \begin{center}
    \includegraphics[scale=1.2]{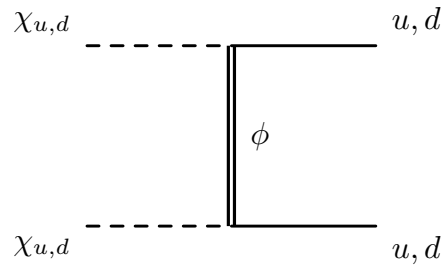}
  \end{center}
  \caption{Diagram contributing to direct detection for dark matter
  coupling to first generation quarks.}
  \label{fig:ddUU}
\end{figure}
MFV suggests that the lightest state in the dark matter multiplet
carries either the flavor of a first generation quark, or a third
generation quark. The direct detection signals are very different in the
two cases, and so we consider them separately. For concreteness we begin
by assuming that the different flavors of $\chi$ are not degenerate, and
that a single dark matter flavor constitutes all the observed dark
matter. We will relax this assumption later.

If dark matter carries up or down flavor, it can scatter off quarks in
the nucleus at tree level by exchanging the mediator $\phi$ as shown in
Fig.~\ref{fig:ddUU}. Starting from the interaction in
Eq.~\ref{quarkdirect} we can integrate out the field $\phi$, leading to
the effective operator
 \begin{align}
  \frac{\lambda^2}{4 m_\phi^2}
  \bar{\chi} (1+\gamma^5) q \;
  \bar{q} (1-\gamma^5) \chi.
 \end{align}
After Fierz rearrangement, this operator becomes,
 \begin{align}
  \frac{\lambda^2}{8 m_\phi^2}
  \bar{\chi} \gamma^\mu (1-\gamma^5) \chi \;
  \bar{q} \gamma^\mu(1+\gamma^5) q.
\label{fierzquark}
 \end{align}
\begin{figure*}[htp]
    \subfloat[]{
    \psfrag{K}[b]{$^{\lambda}/_{m_\phi}$ (GeV$^{-1}$)}
    \psfrag{M}[]{$m_\chi$ (GeV)}
    \includegraphics[scale=0.7]{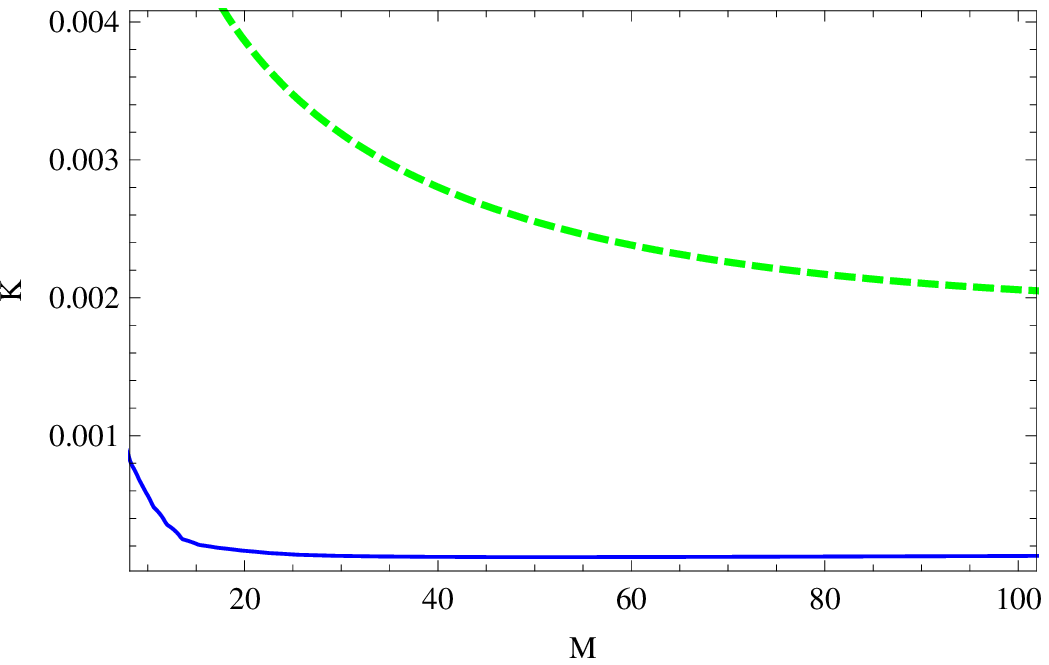}
    }
    \hspace{0.3in}
    \subfloat[]{
    \psfrag{K}[b]{$^{\lambda}/_{m_\phi}$ (GeV$^{-1}$)}
    \psfrag{M}[]{$m_\chi$ (GeV)}
    \includegraphics[scale=0.7]{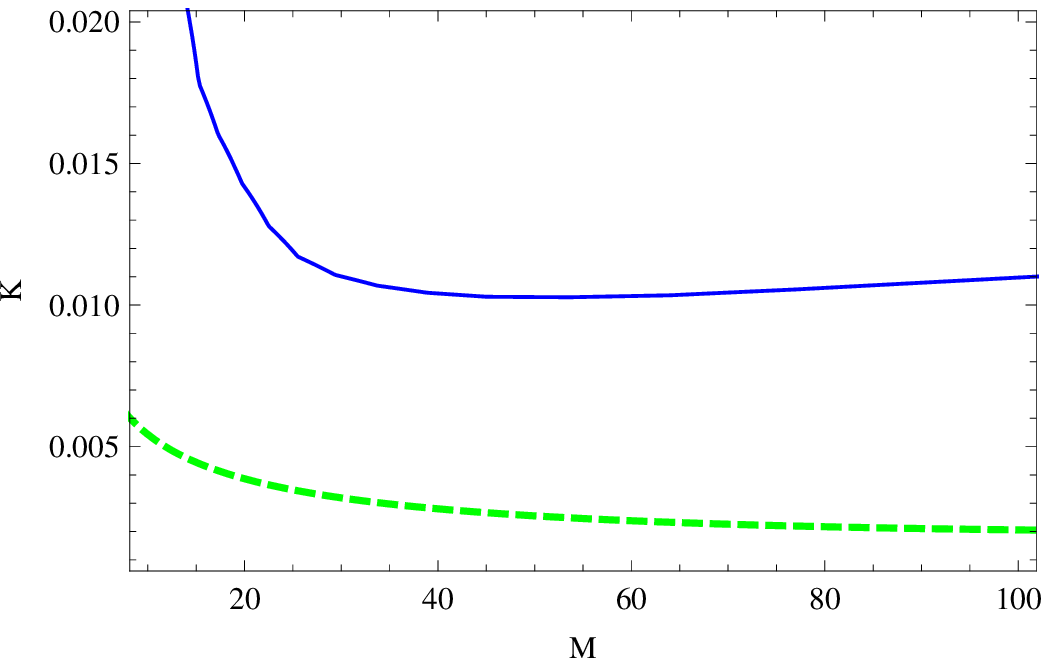}
    }
  \caption{Direct detection and
  relic abundance constraints on quark flavor dark matter for
  a) $\chi_u$ and b) $\chi_b$, when $m_\phi = 150$ GeV. The area above
  the solid blue curve is ruled out by the new
  Xenon100 data~\cite{Aprile:2011hi}. The green dashed
  curves signify the parameters for which we obtain correct relic
  abundance.}
  \label{fig:quark-dd-relic}
\end{figure*}
\begin{figure*}[htp]
  \subfloat[]{
    \includegraphics[scale=0.9]{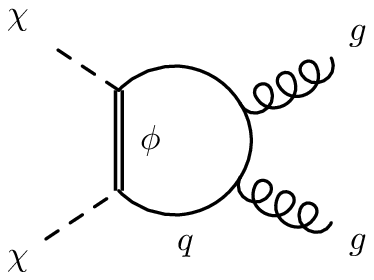}
    \quad
    \includegraphics[scale=0.9]{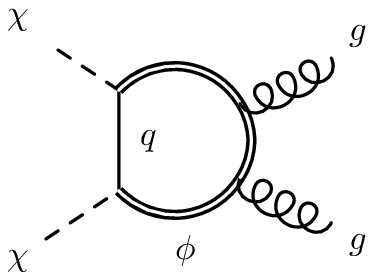}
    \quad
    \includegraphics[scale=0.9]{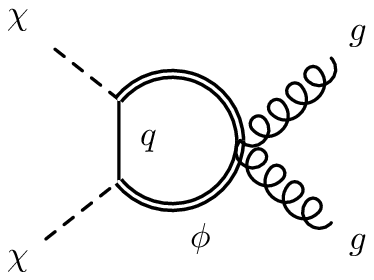}
    \quad
    \includegraphics[scale=0.9]{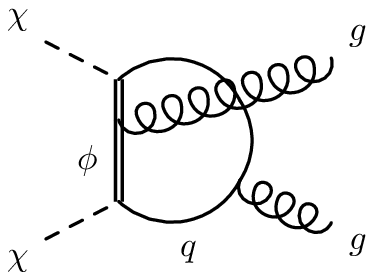}
    }
    \\
    \subfloat[]{
    \includegraphics[scale=1.2]{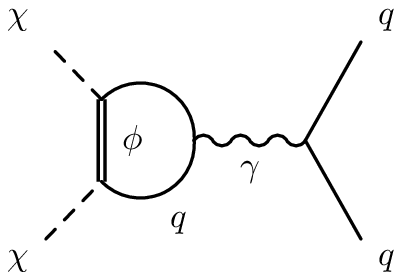}
    \qquad\qquad
    \includegraphics[scale=1.2]{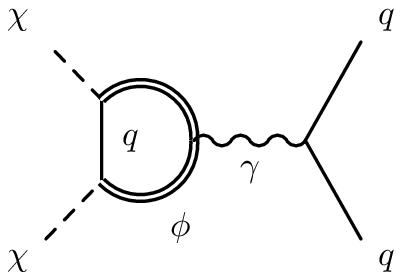}
    }
  \caption{Diagrams contributing to direct detection for dark matter
  coupling to third generation quarks. The scattering can be off a)
  gluons and b) quarks via photon exchange. As discussed in the text,
  the photon exchange dominates for the example considered.}
  \label{fig:ddQQ}
\end{figure*}
The dominant contribution to direct detection comes from the
spin-independent vector-vector coupling. The dark matter-nucleus
cross section (at zero momentum transfer) in this case is given
by~\cite{Agrawal:2010fh}
\begin{align}
  \sigma_0 &=\frac{\mu^2 \lambda^4}{64\pi m_\phi^4}
  \left[A+Z\right]^2
  \\
  \sigma_0 &=\frac{\mu^2 \lambda^4}{64\pi m_\phi^4}
  \left[2A-Z\right]^2,
 \end{align}
for dark matter coupling to up-type and down-type quarks respectively.
Here $\mu$ represents the reduced mass of the dark matter-nucleus
system, while $Z$ and $A$ are the atomic number and mass number of the
nucleus. For a given value of $\lambda$, this cross section is much
larger than in the leptonic case. In fact, as shown in
Fig.~\ref{fig:quark-dd-relic}, the region of parameter space where $\chi$
can be a thermal relic is already excluded by direct detection experiments.

We now move on to the case where the dark matter carries the flavor
quantum numbers of third generation quarks. The contribution arising
from Fig.~\ref{fig:ddUU} is now suppressed by mixing angles, and is
expected to be sub-dominant. There is a possible contribution to the
cross section arising from the one loop diagrams shown in Fig.
\ref{fig:ddQQ}(a), where $\chi$ scatters off gluons in the nucleus.
However, it turns out that this is also not a significant effect. To
understand why, we again integrate out the mediator $\phi$ at tree
level to obtain the effective
operator shown in Eq.~\ref{fierzquark}. This operator allows dark
matter to scatter off a pair of gluons through triangle diagrams
involving quarks.  In general both the vector and axial vector
terms in Eq.~\ref{fierzquark} contribute to the cross section.
However, the contribution from the vector term vanishes identically as
a consequence of the charge conjugation symmetries of QCD and QED
(Furry's theorem)\cite{Kaplan:1988ku,Ji:2006vx}.  The axial vector
interaction couples dark matter to gluonic operators that are parity
odd rather than parity even~\cite{Kaplan:1988ku,Fan:2010gt}. The
parity symmetry of QCD can be
used to show that the matrix elements of these operators in the
nucleus are either spin-dependent or velocity suppressed in the
non-relativistic limit, and therefore do not
contribute significantly to dark matter scattering.

Therefore, the dominant contribution to the cross section arises from
one loop diagrams of the same form as in the lepton case, but now with
the quarks running in the loop (Fig.~\ref{fig:ddQQ} (b)).  The cross
sections will be identical except for factors of color and charge. As
before, we only use the charge-charge interaction to calculate the
bounds.
\begin{align}
  \sigma^0_{ZZ}
  &= \frac{\mu^2 Z^2}{\pi}
  \left[
  \frac{3\lambda^2 e^2 Q}{64\pi^2 m_\phi^2}
  \left[1+\frac23\log\left(\tfrac{m_q^2}{m_\phi^2}\right)
  \right]
  \right]^2,
\end{align}
where $Q=\frac23, -\frac13$ for top and bottom quarks respectively, and
$m_q$ is the mass of the quark in the loop. We see from
Fig.~\ref{fig:quark-dd-relic} that the cross section corresponding to a
thermal WIMP is within reach of current direct detection experiments.

Our MFV analysis shows that if the lightest state carries the up or down 
flavor, the mass splitting between the lightest two states in the dark 
matter multiplet is small enough that the next to lightest flavor plays 
a role in the relic abundance calculation. It may also be stable on 
cosmological time scales, and contribute to the observed dark matter. It 
is therefore important to take this effect into account. Depending on 
the parameters, the dark matter flavors associated with the third 
generation may also play a role in determining the relic abundance. The 
analysis of the previous section shows that if dark matter transform 
under U(3$)_D$, the range of parameters that give rise to the correct 
relic abundance does not differ significantly from the single flavor 
case. It follows that if the lightest flavor is $\chi_d$, this scenario 
is excluded, just as in the single flavor case. If $\chi_b$ is the 
lightest flavor, the other states will decay to it provided it is split 
from them by a few hundred MeV or more. The direct detection bound on 
$\chi_b$ is then the same as in the case when the flavors are 
non-degenerate.

If dark matter transforms under U(3$)_Q$, the couplings of $\chi$ are
hierarchical, and so if the lightest state is associated with the first
generation, it must be quasi-degenerate with the others to obtain the
correct relic abundance. The analysis of the previous section then shows
that if the splittings between these states are small compared to the
freeze out temperature the coupling $\lambda$ corresponding to the third
generation must be a larger by a factor of $\sqrt{3}$ relative to the
single flavor case. If the lightest state, whether belonging to the
first generation or third generation, is split from the other flavors by
more than a few hundred MeV, the observed dark matter today will be
composed of a single flavor. It follows that in this scenario, if the
observed dark matter is composed of third generation particles, the
direct detection limit on $\lambda$ is stronger by a factor of
$\sqrt{3}$ than in the single flavor case. On the other hand, if dark
matter today is composed of particles associated with the first
generation, the limits are much weaker than in the corresponding single
flavor case because of the hierarchical couplings of $\chi$. If more
than one flavor constitutes the observed dark matter, the bound depends
on the constituent fraction of third generation particles and may be
obtained by interpolation. These considerations can be extended in a
straightforward way to the other realizations of MFV.

If dark matter transforms non-trivially under the SM SU(2) symmetry,
scattering processes via $Z$-boson exchange can give large direct
detection signals. Consequently, these scenarios are expected to be
severely constrained. However, if $\chi$ arises from a representation
which does not transform under hypercharge, then it does not couple
directly to the $Z$, and so this effect does not arise. Then the effects
of loop diagrams involving $W$ bosons~\cite{Cirelli:2005uq} must be
compared against the contribution above in order to determine the
leading effect.

\subsubsection{Collider Signals}

The collider signals of this class of theories differ depending on
whether the dark matter particle couples primarily to the third
generation quarks, or to the quarks of the first two generations. We
will restrict our discussion to the case where the couplings of $\chi$
are consistent with MFV. Our results can be extended to the more
general case without difficulty. The mediators $\phi$ can be
pair-produced at the LHC through QCD, and each will
decay down to the dark matter particle either directly, or through a
cascade. If the dark matter particle belongs to the third generation,
at the partonic level each event is associated with two heavy flavor
quarks and up to four light quarks. If, on the other hand, it belongs
to the first generation, we expect between zero and four
heavy flavor quarks in each event, along with two light quarks.

\subsection{Dark Matter with Internal Flavor}

Finally we consider the possibility that dark matter carries a new
internal flavor quantum number that is distinct from either quark or
lepton flavor, and does not couple directly to the SM matter fields at
the renormalizable level. In this framework, the only possible direct
interactions of $\chi$ with the SM fields at the renormalizable level
are to the weak gauge bosons or to the Higgs as shown in Fig.
\ref{fig:dmsmint}(b). These interactions do not generate large new
sources of quark or lepton flavor violation. The direct detection signals
are very similar to those of the corresponding theory where dark matter
does not carry flavor.

These theories are closely related to those where new particles, such as
a scalar boson $\phi$ or vector boson $Z'$, that have couplings of
exactly the same form as in \ref{fig:dmsmint}(b), mediate interactions
between the SM matter fields and the dark matter sector. One important
difference is that these can potentially give rise to SM flavor
violating effects, if their couplings to the SM fields are off-diagonal.

In this scenario, the dark matter states corresponding to different
flavors may be exactly degenerate, if the internal flavor symmetry is
exact, or split, if the symmetry is broken. The collider phenomenology
is highly sensitive to both the splitting between states, and to the
particles produced when heavier states decay to lighter ones. The
heavier particles in the dark matter multiplet can be pair produced
through their couplings to the $Z$, the Higgs, $\phi$ or $Z'$, and can
then decay down to the lightest state. The additional particles produced
in these decays, if visible, together with missing energy, constitute
the collider signatures. This is a natural framework for a hidden
valley~\cite{Strassler:2006im} where particles such as $\phi$ or $Z'$
are the portal to the hidden sector. In this scenario, decays may be
slow on collider time scales, giving rise to displaced vertices, since
the couplings involved can be small in a technically natural way.

\section{Collider Signals of Tau Flavored Dark Matter}

In this section we discuss in detail the collider signals of a specific
model in which the dark matter particle carries quantum numbers under
tau flavor.
For concreteness we assume that  the dark matter particle is a Dirac fermion which is a singlet under weak
interactions, and therefore does not transform under the SU(2) gauge
interactions of the SM.

Dark matter couples directly to the SM through interactions of the form
\begin{align}
  \mathcal{L}
  &=
  \sum_{i=e,\mu,\tau}
  \left[
  \lambda^{i}_{j} \,{E^c}_i \chi^j \, \phi
   \; + {\rm h.c.}
  \right].
\end{align}
where $\phi$ is the mediator, and $\chi_{e,\mu,\tau}$ are the dark
matter and its copies. This interaction fixes the SM quantum numbers of
$\phi$, which is charged under the photon and the $Z$, but does not couple
to the $W$.

For concreteness, we consider two benchmark spectra that are
consistent with MFV, with $\chi$ transforming under U(3)$_E$. Then $\chi_e$ and $\chi_{\mu}$ are expected to be
nearly degenerate since the corresponding SM Yukawa couplings are very
small. We assume that $\chi_\tau$ is lighter than $\chi_e$ or
$\chi_{\mu}$, and constitutes dark matter.

We label the first benchmark spectrum $\tau$FDM1,
\begin{align}
  \label{eq:fdm-spectrum}
  m_{\chi,e} &= 110 \mathrm{\ GeV}\\\nonumber
  m_{\chi,\mu} &= 110 \mathrm{\ GeV}\\\nonumber
  m_{\chi,\tau} &= 90 \mathrm{\ GeV}\\\nonumber
  m_{\phi} &= 160 \mathrm{\ GeV}
\end{align}

The second benchmark spectrum we study has a lighter mediator, and
therefore leads to a larger production cross section (see Fig
\ref{fig:production}). We label this benchmark spectrum $\tau$FDM2:
\begin{align}
  \label{eq:fdm-lite-spectrum}
  m_{\chi,e} &= 90 \mathrm{\ GeV}\\\nonumber
  m_{\chi,\mu} &= 90 \mathrm{\ GeV}\\\nonumber
  m_{\chi,\tau} &= 70 \mathrm{\ GeV}\\\nonumber
  m_{\phi} &= 150 \mathrm{\ GeV}
\end{align}

In these simple models, only the mediator $\phi$ carries SM gauge
quantum numbers, so dark matter events at colliders must arise from
$\phi^+ \phi^-$ production. The $\phi$ particles then decay, either
directly or via cascade decays, into SM charged leptons and the dark
matter particle. Therefore, the characteristic signature of this model
is leptons+MET.

As discussed earlier, MFV restricts the matrix $\lambda$ to be
approximately proportional to identity. Consequently, we take the
couplings of different dark matter flavors to SM to be equal, their
common value set by the relic abundance requirement. Collider signals
are insensitive to this value.

\begin{figure}[tp]
  \psfrag{M}[]{$m_\phi$ (GeV)}
  \psfrag{sig}[]{$\sigma_{pp\rightarrow\phi\phi}$ (fb)}
    \includegraphics[scale=0.75]{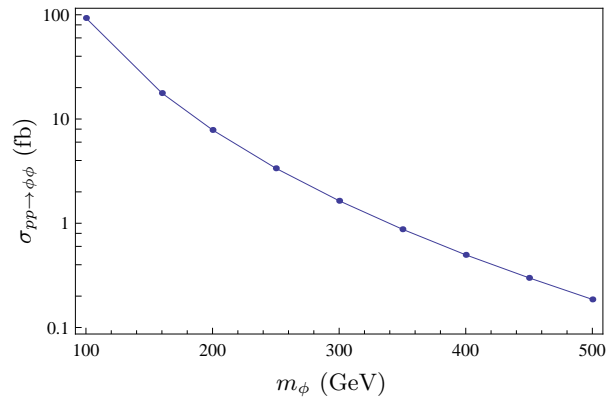}
  \caption{Pair-production cross section for the mediator $\phi$ at
  the 14~TeV LHC run as a function of its mass.}
  \label{fig:production}
\end{figure}
Since $\phi^+ \phi^-$ production proceeds through Drell-Yan and $\phi$
is a scalar, the $\phi$ pair comes out in a $p$-wave, leading to a
small cross section, on the order of $10$~fb at the 14~TeV LHC run.
Therefore, we do not expect the early LHC data to be able to probe
this model. In order to obtain reasonable signal over background
discrimination, tens of inverse fb of data will be required.

\subsection{Signal topologies}

Signal events come in three distinct topologies (see
Fig.~\ref{fig:topology}). Each $\phi$ can decay directly into the dark
matter particle and a $\tau$, corresponding to a short chain.
Alternatively, it can decay to one of the heavier particles in the
dark matter multiplet, which eventually cascades down to the dark
matter particle, creating a long chain. Therefore each event can be
categorized as comprising of short-short, short-long or long-long
chains.

Since $\tau$'s are difficult to identify, we implicitly restrict
ourselves to $\ell=e,\mu$ final states in this section when we talk
about leptons. The events with the long-long decay chain topology have
four-lepton final states (not to mention a pair of $\tau$'s), which
have small SM backgrounds. When $\chi$ is Majorana rather
than Dirac, the short-long chain will also include a like-sign
dilepton final state which is a very clean signal. The $\tau$'s in the
event could also decay leptonically, giving rise to additional
leptons. However, these leptons are generally softer than the primary
leptons. We focus on the long-long decay
topology as the most promising  channel.

In order to simulate signal and background events we use the usrmod
utility of MadGraph/MadEvent \cite{Maltoni:2002qb,Alwall:2007st}, and
we use BRIDGE \cite{Meade:2007js} for the $\chi_{e,\mu}$ decays.
Pythia~\cite{Sjostrand:2006za} is used to simulate parton showers and
hadronic physics, and PGS \cite{PGS} with the default CMS parameter
set is used to simulate detector effects.

\subsection{Backgrounds}

While four-lepton final states are rare in the SM, the signal cross
section is also small so we carefully consider the three leading
sources of backgrounds and devise cuts to reduce them as much as
possible.

\subsubsection{$(Z/\gamma)^{(*)}(Z/\gamma)^{(*)}$}

One of the dominant backgrounds is production of two opposite-sign,
same-flavor lepton pairs from either on-shell or off-shell $Z$'s and
photons. Any missing energy in this background arises from
mis-measurement of lepton momenta, which is small. For the following
contributions to this background, we choose the following cuts:
\begin{itemize}
  \item
    $Z\rightarrow \ell^+ \ell^-$: This is the dominant component in
    this background, which we reduce by imposing a $Z$-veto (described
    in the next subsection)
  \item $Z\rightarrow
    \tau^+\tau^-\rightarrow \ell^+\ell^-$: Even though the Z is
    on-shell in this process, the $Z$-veto is not effective due to the
    presence of neutrinos in the final state. This contribution is
    small due to the leptonic $\tau$ branching ratios. The leptons
    arising from $\tau$ decays are also softer, which we reduce by
    demanding the leptons to be energetic.
  \item $Z^*/\gamma^* \rightarrow \ell^+\ell^-$: While the off-shell
    production cross section is much smaller than on-shell production,
    this contribution is the main one that remains after the $Z$-veto
    and lepton energy cuts. We impose a missing energy cut to reduce
    this background component.
\end{itemize}
\subsubsection{$t\bar{t}(Z/\gamma)^{(*)}$ }

This background process, while it has a three-body final state, has a
cross section comparable to the above process which is purely
electroweak. When both tops decay leptonically and the
$(Z/\gamma)^{(*)}$ goes to leptons, the final state is
$4\ell+$jets+MET. The $Z$-veto reduces the on-shell $Z$ production,
and we also impose a dijet veto (described in the next subsection) in
order to reduce this background, since signal events will typically
not have any hard jets.

\subsubsection{$WW(Z/\gamma)^{(*)}$ }

This process is qualitatively similar to the above process, but has a
much smaller production cross section because it is purely
electroweak. On the other hand, there are no additional hard jets in
these events, so they escape the dijet veto. Consequently, events
which escape the $Z$-veto can fake the four-lepton signal very well.
Demanding the leptons to be energetic and imposing the missing energy
cut helps reduce this background.

\subsubsection{Backgrounds with fakes} There are also backgrounds
arising from jets that are misidentified as leptons. We find that
provided the fake rates are of order $10^{-3}$ or less, the irreducible
backgrounds described above are the dominant ones.

\begin{figure*}[htp]
  \subfloat[]{
  \includegraphics[scale=0.6]{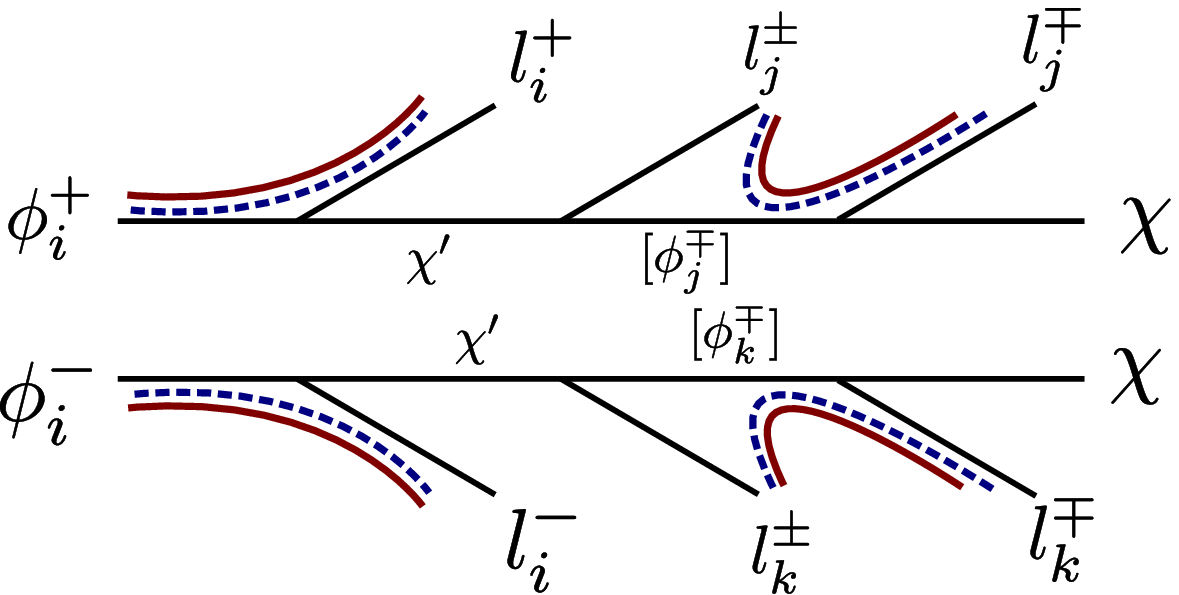} }
  \hspace{0.2in}
  \subfloat[]{ \includegraphics[scale=0.6]{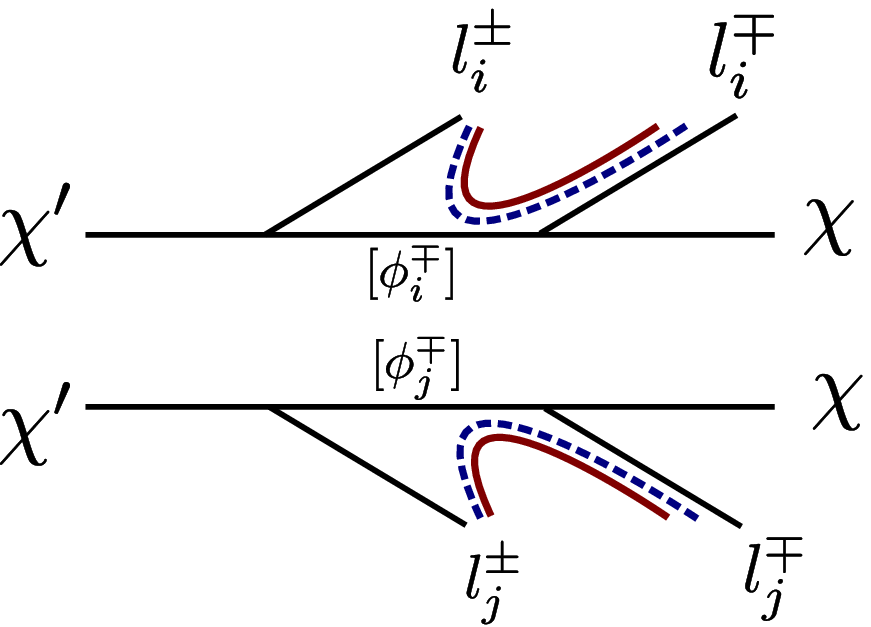} }
  \caption{Flavor (red solid) and charge (blue dashed) correlations
  are shown for topologies in strawman models.}
  \label{fig:flavor-charge-susy}
\end{figure*}

\begin{figure*}[htp]
  \includegraphics[scale=0.6]{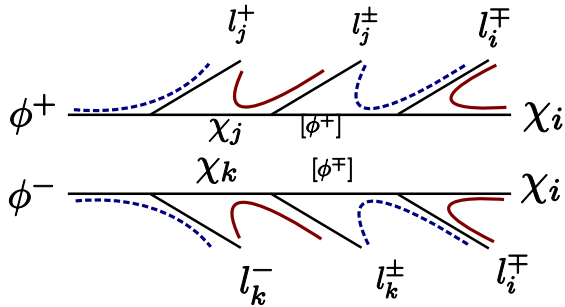}
  \caption{Flavor
  (red solid) and charge (blue dashed) correlations are shown for
  $\tau$FDM. Final state lepton charge ambiguities for Majorana dark
  matter models do not affect the charge correlation.}
  \label{fig:flavor-charge-fdm}
\end{figure*}

\subsection{Cuts}

We use the following cut flow in order to maximize signal over
background: \begin{itemize} \item{\underline{Lepton cuts}} - We
demand events with at least four leptons each with $pT > 7$ GeV. At
least two of these leptons are further required to have $E> 50$ GeV.
\item{\underline{Dijet veto}} - We discard events with two or more
jets of $pT>30$ GeV each.  \item{\underline{$Z$ veto}} - We veto
events if the invariant mass of any $Z$-candidate (a pair of
same-flavor and opposite-charge leptons) falls within 7 GeV of the
$Z$ mass. This is a tighter $Z$-veto than is usually used, but we
find that the loss in signal efficiency is more than compensated for
by the background reduction.  \item{\underline{Missing energy}} - We
require at least 20 GeV of missing energy in each signal event. Since
most backgrounds with high MET have already been eliminated by the
previous cuts in the cut flow, we find that a mild threshold such as
20 GeV is sufficient.  \end{itemize}

\subsection{Results}

\begin{table}[tp] \vspace{0.3in}
  \begin{center}
    \renewcommand{\tabcolsep}{5pt}
    \begin{tabular}{ lrrrr
      }
      \hline \hline
      \multirow{2}{*}{Dataset}& \multicolumn{4}{c}{Event rate after cuts
      at 100 fb$^{-1}$} \\
      &Lepton cuts&Jet cuts&$Z$ veto&MET
      \\ \hline
      $\tau$FDM1 & 46.73& 42.83& 38.41& 35.01\\
      $\tau$FDM2 &
      75.39 & 69.30& 63.26&57.04\\
      $\ell^+ \ell^- \ell^+\ell^- $&1617.94&1582.42&140.30&13.32\\
      $t\bar{t} \,\ell^+\ell^- $&89.57&19.45&4.92&4.70 \\
      $WW\ell^+\ell^- $&14.70&13.98&2.51&2.51 \\
      \hline
      \hline
    \end{tabular}
  \end{center}
  \caption{Signal and SM background event rates for
  processes yielding 4-lepton  final states after each set of
  cuts is progressively applied (note that $\ell=e,\mu$).  All numbers
  are reported for the 14 TeV LHC run and include detector effects.}
  \label{tab:SMbg}
\end{table}

  The signal and background events of each type that
survive these cuts are listed in Table~\ref{tab:SMbg}. These results
show that it is possible to discover the $\tau$FDM2 benchmark above
SM backgrounds at 5$\sigma$ significance with about 20 fb$^{-1}$ of
data at the 14 TeV LHC run. A higher luminosity ($\sim 40$ fb$^{-1}$)
would be needed in order to distinguish the $\tau$FDM1 benchmark from
the SM background.  Note that while we have based this expectation on
statistical uncertainties only, we have been conservative in many
other aspects.  In particular, a requirement that each event have at
least one $\tau$ candidate would virtually eliminate all remaining
backgrounds while reducing the signal only moderately. Furthermore,
one could do better than a pure counting experiment by taking into
account the charge and flavor correlations present in the signal,
which are different than the backgrounds in order to further increase
sensitivity. We will indeed use this approach in the next section
where we consider how the FDM model could be distinguished from more
conventional DM models where the DM particle is a flavor singlet.

While ATLAS~\cite{ATLAS-CONF-2011-039} and
CMS~\cite{Chatrchyan:2011ff} have already performed searches in
multilepton final states, considering the low cross section of the FDM
benchmark model, they are not yet expected to have exclusion level
sensitivity to this scenario.

\section{Distinguishing $\tau$FDM}

\begin{figure*}[htp]
  \begin{center}
    \psfrag{FDM1}[][0.4]{$\tau$FDM1}
    \psfrag{S1}[]{Spectrum 1}
    \psfrag{S2}[]{Spectrum 2}
    \psfrag{S3}[]{Spectrum 3}
    \psfrag{M}[]{Mass (GeV)}
    \psfrag{phi}{$\phi$}
    \psfrag{chitau}{$\chi_\tau$}
    \psfrag{chiemu}{$\chi_{e,\mu}$}
    \psfrag{chi}{$\chi$}
    \psfrag{chip}{$\chi'$}
    \psfrag{sleptons}{$\widetilde{e},\widetilde{\mu},\widetilde{\tau}$}
    \includegraphics[width=0.9\textwidth]{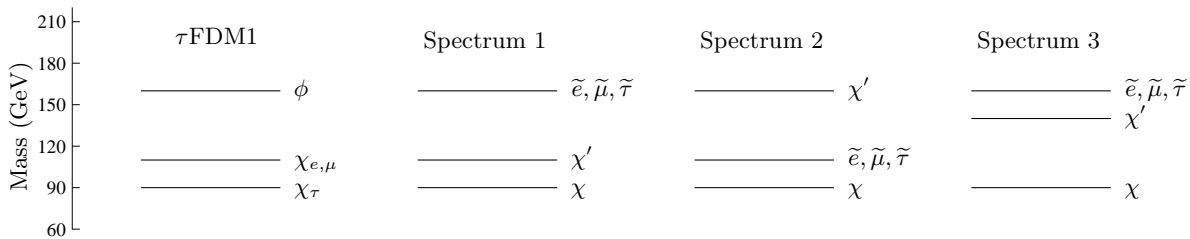}
  \end{center} \caption{The FDM spectrum and the strawman spectra
  compared}
  \label{fig:spectra}
\end{figure*}

Multi-lepton events with large missing energy are fairly common
signals in theories with neutral stable particles and partners to the
SM leptons, which include a variety of dark matter models. We would
to like understand whether it is possible to distinguish at the LHC
the model of $\tau$FDM that we studied in the previous section from
models with similar signatures but where the dark matter does not
carry flavor quantum numbers. Clearly, this question is very
difficult in general. Therefore, we focus on a more restricted
question. We investigate whether it is possible to distinguish
$\tau$FDM from a specific `strawman' model, where the dark matter
does not carry flavor.

The strawman model we choose is related to supersymmetric theories
where the bino constitutes dark matter. The form of the
lepton-slepton-bino vertex is very similar to the defining vertex of
a theory of lepton FDM, except that in the supersymmetric case it is
the slepton that carries flavor, not the bino. The strawman model we
choose therefore consists of the bino, which we label by $\chi$,
along with the three right-handed sleptons, ${\widetilde{E}}^c_i$.
The bino constitutes dark matter. To mimic the collider signals of
$\tau$FDM, we add to the strawman model an additional `neutralino'
$\chi'$, which is heavier than the bino. $\chi'$ is an admixture of a
SM SU(2) doublet and singlet, so that it can be pair-produced through
the $Z$, and is chosen to couple to leptons and sleptons in a
flavor-blind way.  This interaction takes the schematic form
\begin{align} \lambda' {E_i^c} \chi' {\widetilde{E}^{c \; i}}  \; \;
+ \; \; {\rm h.c.} \end{align} The couplings of $\chi'$ are somewhat
different from those of a conventional neutralino in the MSSM, since
any neutralino with significant couplings to the $Z$ is expected to
contain a significant Higgsino component, and the Higgsino does not
couple universally to the different leptons.  However, this simple
strawman model captures the main features of theories where the dark
matter does not carry flavor, while generating events which are very
similar to those of $\tau$FDM.

For simplicity, in what follows we assume that the three sleptons are
degenerate in mass. In general, $\chi'$ could either be lighter than
or heavier than the sleptons, while the bino is the lightest of the
new states. Both $\chi'$ and $\chi$ are taken to be Majorana fermions
as is the case in the MSSM.

Signal events in the $\tau$FDM model involve four or more isolated
leptons and missing energy. How does the strawman model generate
similar events? The sleptons can be pair-produced in colliders. If
they are heavier than $\chi'$, this leads to events of the form shown
in Fig.~\ref{fig:flavor-charge-susy}(a), which involve six leptons,
any or all of which could be taus. We label this possibility topology
(a). Two $\chi'$ particles can also be pair-produced, leading to
events of the form shown in Fig.~\ref{fig:flavor-charge-susy}(b),
which we label topology~(b). These events involve four leptons, any or all
of which could be taus.

How can we distinguish between signal events in the two classes of
models? One possibility is to note that we expect exactly two taus in
each signal event in the $\tau$FDM model, whereas events in the
strawman models will involve between zero and six. This could be a
useful discriminant as the LHC experiments continue to improve their
$\tau$ identification capabilities. Presently, we do not make use of
this discriminant. We also do not assume that the total event rate
will be a reliable variable for discrimination. Even though in
principle the event rate can vary widely across different models, for
the following analysis we simply scale the event rate from the
strawman model to match the number of events from $\tau$FDM and
concentrate on ratios and asymmetries.

In particular, we focus on charge and flavor correlations among the
final state leptons in the event. In
Fig.~\ref{fig:flavor-charge-susy} and
Fig.~\ref{fig:flavor-charge-fdm}, we exhibit the correlation of
flavor and charge among the final state leptons in signal events for
the $\tau$FDM model and the strawman model. The crucial observation
is that, in the case of $\tau$FDM, for the chosen spectrum, the two
upstream leptons are also the hardest. This is likely to be the case
for spectra motivated by MFV. These leptons are charge
anti-correlated since they arise from the decay of the charged
mediators $\phi$.  However, they have no flavor correlation, because
the mediator does not carry flavor quantum numbers.

Contrast this with the strawman model. Consider first events
associated with topology (a). If the mass of $\phi$ is much larger
than that of $\chi'$, the two upstream leptons in the event are the
hardest. These exhibit charge anti-correlation, but are flavor
correlated, in contrast to $\tau$FDM. If, on the other hand, the mass
of $\phi$ is close to that of $\chi'$, two of the four downstream
leptons will be the hardest. However, these exhibit no significant
charge or flavor correlation, unlike $\tau$FDM. What about events
associated with topology (b)? Here the two hardest leptons are again
charge and flavor uncorrelated. We conclude from this that the charge
and flavor correlations are different in the two theories, and may
allow them to be distinguished.

We generated signal events for the strawman model for three benchmark
spectra, shown schematically in Fig.~\ref{fig:spectra}, and compared
the resulting charge and flavor correlations to those of $\tau$FDM.
In particular, the spectra we studied were the following.

\smallskip \noindent \underline{Spectrum 1} \\
We assume the masses
of the sleptons to be the same as that of the mediator $\phi$ in
$\tau$FDM. The masses of $\chi'$ and $\chi$ are also chosen equal to
the $\chi_{e,\mu}$ and $\chi_\tau$ mass respectively. Then,
\begin{align}
  \label{eq:susy-spectrum1}
  m_{\chi'} &= 110 \mathrm{\ GeV}\\
  \nonumber m_{\chi} &= 90 \mathrm{\ GeV}\\\nonumber
  m_{\widetilde e, \widetilde \mu, \widetilde\tau}
  &= 160 \mathrm{\ GeV}
\end{align}

This spectrum can clearly give rise to both topologies in
Fig.~\ref{fig:flavor-charge-susy}, but since the mass splitting
between $\chi$ and $\chi'$ is small, events from topology (b)
generally fail to pass the four-lepton cut requiring two leptons to
have more than 50 GeV energy. Therefore, topology (a) dominates the
phenomenology of this benchmark.

In this topology, the two most upstream leptons are also the hardest,
{\it and} are flavor-correlated. The $\tau$FDM leptons, as noted
above, have no flavor correlation. Therefore, we expect that the
flavor-correlation of the two hardest leptons is a good discriminant
in this case.

\smallskip \noindent \underline{Spectrum 2} \\
If the mass of the
sleptons is less than the mass of $\chi'$, then only the topology (b)
is allowed. The decay of $\chi'$ is on-shell in this case.

The representative spectrum we study is,
\begin{align}
  \label{eq:susy-spectrum2}
  m_{\chi'} &= 160 \mathrm{\ GeV}\\
  \nonumber m_{\chi} &= 90 \mathrm{\ GeV}\\
  \nonumber m_{\widetilde e, \widetilde \mu, \widetilde\tau}
  &=110 \mathrm{\ GeV}
\end{align}
In this case,
the hardest leptons should exhibit neither charge nor flavor
correlations, allowing us to distinguish it from $\tau$FDM.

\smallskip \noindent \underline{Spectrum 3} \\
Consider again the case
when the mass of $\chi'$ is less than mass of $\chi$.  As noted in the
case of Spectrum 1, when the mass of $\chi'$ is close to the mass of
$\chi$, topology (a) dominates.  On the other hand, when the mass of
the $\chi'$ is very close to the mass of sleptons, the most upstream
leptons become softer, and topology (b) dominates.  The
conclusions in this case are then identical to those of Spectrum 2.

In the
intermediate case, however, the result is a mixture of the two
topologies. In
order to investigate this we study a third spectrum,
\begin{align}
  \label{eq:susy-spectrum3}
  m_{\chi'} &= 140 \mathrm{\ GeV}\\
  \nonumber
  m_{\chi} &= 90 \mathrm{\ GeV}\\
  \nonumber m_{\widetilde e, \widetilde
  \mu, \widetilde\tau} &= 160 \mathrm{\ GeV}
\end{align}
In the next section we study the extent to which each of these spectra can be
distinguished from $\tau$FDM.

\subsection{Comparison}

The correlations we obtain are
listed in Table~\ref{tab:FDMvsSUSY}.  The results are in agreement
with our expectations. Events with topology
(a) in Spectrum 1 clearly exhibit flavor correlation between the two
hardest leptons, as expected for the upstream leptons created from
(flavor-carrying) sleptons.  $\tau$FDM, on the other hand exhibits no
flavor correlation in the hardest two leptons.

\begin{table}[htp]
  \begin{center}
    \begin{tabular}{
      m{0.9in}>{\centering}
      m{0.6in}
      m{0.6in}<{\centering} }
      \hline \hline
      \multirow{2}{*}{Dataset}
      &\multicolumn{2}{c}{Frac. events with same}\\
      &Flavor&Charge\\ \hline
      $\tau$FDM1    &0.52&0.14\\
      $\tau$FDM2    & 0.49&0.14\\
      Spectrum 1(a) & 0.87&0.13\\
      Spectrum 1(b) & 0.61&0.39\\
      Spectrum 2    & 0.55&0.41\\
      Spectrum 3(a) & 0.66&0.33\\
      Spectrum 3(b) & 0.60&0.38\\
      \hline
      \hline
    \end{tabular}
  \end{center}
  \caption{Flavor and charge correlations for the two highest $p_{\rm T}$
  leptons in events passing cuts for different data samples. The
  strawman models are represented by the spectrum and their event
  topology.}
  \label{tab:FDMvsSUSY}
\end{table}

In all the fake spectra with topology (b), the two hardest leptons
show no preferential charge assignment beyond the ratio of $1:2$ for
same to opposite charge, as expected from random charge assignment.
Consequently these cases have a weaker charge anti-correlation than
the $\tau$FDM.

Events from topology (a) in Spectrum 3 fall in the middle, with
somewhat significant charge anti-correlation, and a weak flavor
correlation. While the correlation between charge and flavor is
different from the $\tau$FDM case, higher statistics might be needed
in this case to make a precise distinction. We also illustrate these results in fig.~\ref{fig:flav-charge-plot} where we plot the flavor
and charge asymmetries of the two hardest leptons for FDM and the strawman spectra, taking into account the contributions from SM backgrounds. The charge and flavor asymmetries are defined as
\begin{align} a_F,a_C =
  \frac{n_{\mathrm{same}}-n_{\mathrm{diff.}}}
  {n_{\mathrm{same}}+n_{\mathrm{diff.}}}.
  \label{eq:asymm-def}
\end{align}

In order to assess the effect of statistical fluctuations and to
quantify the distinguishability of FDM from the strawman spectra, we
perform a log-likelihood ratio (LLR) study as follows: We use the
fraction of events where the two hardest leptons have same/opposite
(S/O) sign/flavor (S/F) given in figure \ref{fig:flav-charge-plot}
(obtained from very high statistics Monte Carlo samples) and obtain
for FDM and for each strawman spectrum (SMS) the probability that any
given event will have same/opposite sign/flavor correlations. We
denote these probabilities by $p_{i,\rm FDM}$ for FDM, where $i\in
\{\rm OSOF,OSSF,SSOF,SSSF\}$, and by $p_{i,\rm SMS1a}$ etc. for the
strawman spectra. Given a data sample with $N=N_{\rm OSOF}+N_{\rm
OSSF}+N_{\rm SSOF}+N_{\rm SSSF}$ events, we can compare the likelihood
of the FDM hypothesis to the hypothesis of any one of the strawman
spectra by defining the LLR
\begin{equation}
{\rm LLR} = 2 Log\left(\frac{\prod_i (p_{i,\rm FDM})^{N_i}}{\prod_i (p_{i,\rm SMS})^{N_i}}\right).
\end{equation}

In order to quantify the distinguishability of the FDM model from each of the strawman spectra, we generate a large number of pseudo-data samples by sampling $N_{\rm OSOF}$ etc. around their mean values using a Poisson distribution and fit the distribution of LLR values obtained in this way to a Gaussian. Then, for the comparison of FDM to any one of the strawman spectra, we determine $x_{cut}$, defined as the LLR value where the Gaussian distributions of the two hypotheses intersect. Using $x_{cut}$ we construct a discriminant, such that any data sample with LLR value greater than the $x_{cut}$ is accepted as having arisen from an underlying FDM model, and any data sample with LLR value less than $x_{cut}$ is accepted as having arisen from the corresponding strawman spectrum. The probability for data arising from an FDM model to be "mistagged" as a strawman spectrum, denoted by $p_{\rm FDM\rightarrow SMS}$ is then given by the integral of the FDM LLR distribution between $-\infty$ and $x_{cut}$ and conversely the probability for a data sample arising from a strawman model to be mistagged as FDM, denoted by $p_{\rm SMS\rightarrow FDM}$, is given by the integral of the strawman LLR distribution between $x_{cut}$ and $+\infty$.

Using statistics corresponding to $200~{\rm fb}^{-1}$ of luminosity at the 14 TeV LHC, we then evaluate these mistag rates between FDM and each strawman spectrum. The results are given in table \ref{tab:mistag}. Note that all entries correspond to a confidence-level equivalent to 2$\sigma$ (or higher). In other words, the LLR discrimination described above allows one to distinguish the FDM model from each one of the strawman spectra by at least $95\%$ confidence-level.

\begin{table}[h]
  \begin{center}
    \begin{tabular}{
      m{0.9in}>{\centering}
      m{0.6in}
      m{0.6in}<{\centering} }
      \hline \hline
      Strawman Spectrum & $p_{\rm FDM\rightarrow SMS}$ & $p_{\rm SMS\rightarrow FDM}$\\ \hline
      Spectrum 1(a) & 0.007 & 0.006\\
      Spectrum 1(b) & 0.022 & 0.027\\
      Spectrum 2    & 0.019 & 0.024\\
      Spectrum 3(a) & 0.040 & 0.044\\
      Spectrum 3(b) & 0.026 & 0.032\\
      \hline
      \hline
    \end{tabular}
  \end{center}
  \caption{Based on the log-likelihood analysis described in the text, the probabilities for data arising from an FDM model to be ``mistagged'' as a strawman spectrum, and vice versa.}
  \label{tab:mistag}
\end{table}

\begin{figure}[htp]
  \begin{center}
    \psfrag{FDM1}[][0.4][0.8]{$\tau$FDM1}
    \psfrag{S1a}[][][0.6]{Spectrum 1(a)}
    \psfrag{S1b}[][][0.6]{Spectrum 1(b)}
    \psfrag{S2} [][][0.6]{Spectrum 2}
    \psfrag{S3a}[][][0.6]{Spectrum 3(a)}
    \psfrag{S3b}[][][0.6]{Spectrum 3(b)}
    \psfrag{aC}[]{$a_C$}
    \psfrag{aF}[]{$a_F$}
    \includegraphics[width=0.46\textwidth]{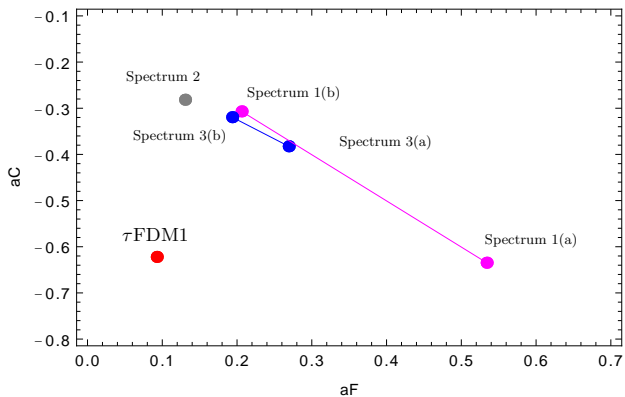}
  \end{center}
  \caption{Flavor and charge asymmetry
  for different models and event topologies. The straight lines interpolate between points which correspond to
  different event topologies for each fake spectrum, in order to account for cases
  where both topologies contribute.}
  \label{fig:flav-charge-plot}
\end{figure}

Note that while we have relied only on the energies of the leptons to
help us identify the topology of the event, this is just the simplest
approach and can be extended with more sophisticated tools. For
example, when applicable, the hemisphere algorithm
\cite{Ball:2007zza} can distinguish particles arising from different
decay chains, and has been used widely for this purpose
\cite{Nojiri:2008hy,Nojiri:2008vq,Agashe:2010tu}. There have also
been other techniques proposed to help identify event topologies
\cite{Rajaraman:2010hy,Bai:2010hd}. And clearly, one can make use of
further ratios involving the softer leptons to obtain better
discrimination between these models.

\section{Conclusions}

In conclusion, we have studied the direct detection and collider
prospects of theories where the dark matter particle carries flavor
quantum numbers, and has renormalizable contact interactions with the
Standard Model fields. We have shown that the phenomenology of this
scenario depends on whether dark matter carries lepton flavor, quark
flavor or its own internal flavor quantum numbers. Each of these
possibilities is associated with a characteristic type of vertex,
leading to different predictions for direct detection experiments and
to distinct collider signatures. In particular, assuming a coupling
consistent with relic abundance considerations, we have shown that
many of these models could be probed in the near future by upcoming
direct detection experiments

We have studied in detail a class of models where dark matter carries
tau flavor, where the collider signals include events with four or
more isolated leptons and missing energy. We have performed a full
simulation of the signal and SM backgrounds, including detector
effects, and shown that in a significant part of the parameter space
favored by MFV, these theories can be discovered above SM backgrounds
at the 14 TeV LHC run. We have also shown that flavor and charge
correlations among the final state leptons may allow models of this
type to be distinguished from simple theories where the dark matter
particle couples to leptons but does not carry flavor.

\begin{acknowledgments}
  We thank Takashi Toma for pointing out an error in an earlier
  version of this manuscript. CK would like to thank Sunil Somalwar and
  Scott Thomas for helpful discussions about details of LHC searches
  and Rouven Essig, Jay Wacker and Eder Izaguirre for their generous
  offer to make their background event samples available to us. PA
  would like to thank Andrzej Buras for useful discussions and
  hospitality at the Technical University, Munich during the
  completion of a part of this work.  We would also like to thank
  Michael Ratz for useful comments.  PA and ZC are supported by NSF
  grant PHY-0801323 and PHY-0968854.  SB acknowledges support from the
  CSIC grant JAE-DOC, as well as from MICINN, Spain, under contracts
  FPA2010-17747 and Consolider-Ingenio CPAN CSD2007-00042. SB is also
  supported by the Comunidad de Madrid through Proyecto HEPHACOS
  ESP-1473. The work of CK is supported by DOE grant DE-FG02-96ER40959
  and by the NSF Grant Number PHY-0969020.
\end{acknowledgments}

\appendix
\section{Direct detection of lepton flavored dark matter}
\label{appendix}
In this section we calculate the contribution to the cross section
for dark matter scattering off a nucleus arising from the diagram
shown in Fig.~\ref{fig:ddleptons}.

\begin{figure*}[htp]
  \begin{center}
    \includegraphics[scale=1.2]{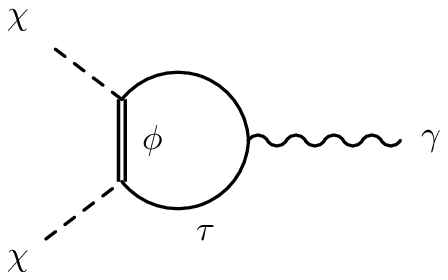} \hspace{0.3in}
    \includegraphics[scale=1.2]{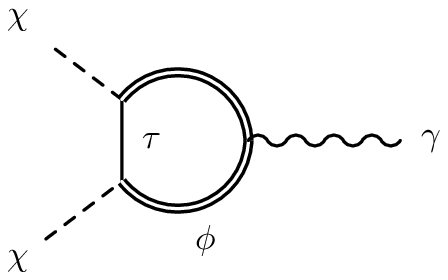} \hspace{0.3in}
  \end{center}
  \caption{Dark matter interaction with a photon in the full theory}
  \label{fig:ddloop}
\end{figure*}

Our approach will be to integrate out the mediator $\phi$ and the
leptons $l$ in the loop to obtain an effective vertex for the
coupling of the dark matter particle $\chi$ to the photon. The
resulting effective theory, which can be used to directly obtain the
cross section, is valid provided the momentum transfer $|\vec{k}|$ in
the process is smaller than the mass of the lepton in the loop. The
momentum transfer in direct detection experiments is typically of
order 10 - 50 MeV, which implies that this procedure is valid if the
lepton in the loop is the muon or the tau, but not if it is the
electron. However, the final result can generalized to obtain an
expression that is approximately valid for this case as well.

We first identify the operators that can potentially appear in the
effective vertex. We begin by noting that the vertices in the
diagrams in Fig.~\ref{fig:ddloop} do not by themselves violate $CP$
symmetry.  We therefore write down the leading effective operators
that couple dark matter to the photon, and which are consistent with
electromagnetic gauge invariance and $CP$. The lowest dimension
operator consistent with these symmetries is unique. It is the
dimension-5 dipole moment operator,
\begin{align} \bar{\chi}
  \sigma_{\mu\nu} \chi \; F^{\mu\nu}
\end{align}
However, this operator
does not actually appear in the effective theory. The underlying
reason is that this operator breaks the chiral symmetry of the $\chi$
field. However, all the vertices and propagators in
Fig.~\ref{fig:ddloop} respect this symmetry, while the mass term of
$\chi$, which breaks it, does not appear in the diagrams. This is
most easily seen if we first integrate out just the heavy mediator
$\phi$ and consider the resultant effective four-fermion operator,
\begin{align}
  \bar{\chi} (1+\gamma_5) \ell \; \bar{\ell}
  (1-\gamma_5)\chi.
\end{align}

A Fierz rearrangement shows that
this is equivalent to the operator
\begin{align} \bar{\chi}
  \gamma^\mu(1-\gamma_5)\chi \; \bar{\ell}\gamma_\mu
  (1+\gamma_5)\ell,
\end{align} which establishes that the dark
matter coupling is indeed chiral.  Hence, we do not generate the
dipole interaction above after
integrating out the lepton.

As a consequence, the leading contribution to the DM-nucleus
scattering arises from dimension-6 operators. Again, gauge and $CP$
symmetries, together with the chiral symmetry mentioned above,
restrict us to the following two operators,
\begin{align}
  \mathcal{O}_1 &= \left[ \bar{\chi} \gamma^\mu(1-\gamma^5)\partial
  ^\nu \chi\; +h.c.\right] F_{\mu\nu} \\ \mathcal{O}_2 &= \left[
  i\bar{\chi} \gamma^\mu(1-\gamma^5)\partial ^\nu \chi\; +h.c.\right]
  F^{\sigma\rho} \epsilon_{\mu\nu\sigma\rho} .
\end{align}
The factors
of $i$ in the definitions of these operators have been chosen so that
their coefficients in the effective theory are necessarily real.

To calculate the coefficients of these operators in the effective
theory we perform a matching calculation from the full theory to the
effective theory, where the mediator $\phi$ and the lepton $l$ have
been integrated out.

In 4-component Dirac notation the relevant part of the Lagrangian in
the full theory takes the form
\begin{align} \mathcal{L} &\supset
  \frac{\lambda}{2} \left[\bar{\chi} (1+\gamma_5)\ell \;\phi
  +\bar{\ell} (1-\gamma_5) \chi\; \phi^\dagger \right]
\end{align}
We can compute the one loop processes shown in Fig.~\ref{fig:ddloop}
to find the low energy effective Lagrangian. Since we are interested
in direct detection processes with momentum transfer of at most
$\mathcal {O}(100$ MeV$)$, we only work to $\mathcal{O}(k^2/m_\phi^2)$
in momentum transfer. Further, we only keep the leading term in
$m_{\ell}/m_\phi$, which is a good approximation in our case. In this
limit, the amplitude in the full theory is given by,
\begin{align}
  \mathcal{M} &=
  \frac{\lambda^2 e }
  {64\pi^2 m_\phi^2} \bar{u}(p_2)
  \gamma_\delta(1-\gamma^5) u(p_1)
  \epsilon^{*}_{\mu}(k)
  \nonumber\\
  &\qquad\qquad\times \left[
  k^2
  \left( \frac 12
  + \frac23\log \left[\tfrac{m_{\ell}^2}{m_\phi^2} \right]
  \right)
  g^{\mu\delta}
  \right.\nonumber\\&\qquad\qquad\qquad\qquad\left.
  -\frac i2 (p_{1}+p_{2})_\alpha k_\beta
  \epsilon^{\alpha\mu\beta\delta} \right],
\end{align}
where $p_1, p_2$ and $k$ are the momenta of the incoming
dark matter, outgoing dark matter, and the photon, respectively.
Using integration by parts on the operators shown above, we see that
the term with $k^2$ corresponds to $\mathcal{O}_1$, and the second
term corresponds to $\mathcal{O}_2$. Matching the coefficients, we
can write down the effective Lagrangian,
\begin{align}
  \mathcal{L}_{eff} &= \frac{-\lambda^2e}{64\pi^2 m_\phi^2} \left[
  \left(
  \frac12
  +\frac23\log\left[\tfrac{m_{\ell}^2}{m_\phi^2}\right]
  \right)
  \mathcal{O}_1 + \frac14 \mathcal{O}_2 \right] .
\end{align}

\begin{figure}[tp]
  \begin{center}
    \includegraphics[scale=1.1]{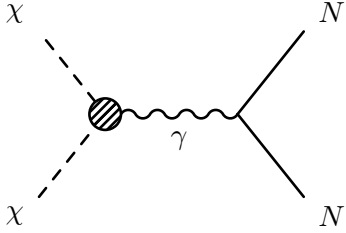}
  \end{center}
  \caption{Direct detection diagram for the dark matter in the effective theory}
  \label{fig:figddNN}
\end{figure}
We can now calculate the amplitudes for scattering of dark matter with
nuclei arising from these different effective operators and
investigate their qualitative behavior.
The scattering amplitude
due to the first operator is given by,
\begin{align}
  \mathcal{M}_{\mathcal{O}_1} &= \sum_q \frac{\lambda^2e^2}{64\pi^2
  m_\phi^2} \left(\frac
  12+\frac23\log\left[\tfrac{m_{\ell}^2}{m_\phi^2}\right]\right)
  \nonumber\\&\qquad\times \bar{u}(p_2) \gamma^\mu(1-\gamma^5)u(p_1) \;
  \langle N| Q\, \bar{q}\gamma_\mu q |N\rangle,
\end{align}
where we
sum the matrix elements of all quark bilinears in the nucleus, and
$Q$ is the charge of the quark in units of $e$.  This is the typical
interaction through the vector current. This gives rise to
predominantly spin-independent cross sections which are enhanced for
large nuclei.

Consider the scattering amplitude due to the second operator,
\begin{align} \mathcal{M}_{\mathcal{O}_2}
  &=
  -\sum_q
  \frac{i\lambda^2e^2}
  {32\pi^2 m_\phi^2}
  \bar{u}(p_2)
  \gamma^\mu(1-\gamma^5)u(p_1) \;
  \nonumber\\&\qquad\times
  \frac{(p_2+p_1)^\nu k^\alpha}{4 k^2}
  \langle N| Q\,
  \bar{q}\gamma^\beta q |N\rangle
  \epsilon_{\mu\nu\alpha\beta} .
\end{align}
To disentangle different contributions, we use the Gordon
identity on the dark matter spinors. Since we are using the equation
of motion of the dark matter particle, we will now generate chiral
symmetry-violating bilinears as well ($\bar{u}\,\sigma^{\alpha\beta} u$
in particular).  Neglecting terms of higher order in momentum
transfer and relative velocity, we get,
\begin{align}
  \mathcal{M}_{\mathcal{O}_2}
  &=
  -\frac{i\lambda^2 e^2}{64\pi^2 m_\phi^2m_\chi}
  \frac{(p_1+p_2)_\omega (p_2+p_1)^\nu k^\alpha}{4k^2}
  \nonumber
  \\&\qquad\qquad\times
  \bar{u}(p_2)
  \left[
  i\sigma^{\omega\mu} \gamma_5
  \right]
  u(p_1)
  \nonumber\\&\qquad\qquad\times
  \sum_q \langle N| Q\,
  \bar{q}\gamma^\beta q |N\rangle
  \epsilon_{\mu\nu\alpha\beta} \; .
\end{align}
We can rewrite $\sigma^{\omega\mu}\gamma^5$ as $\frac i2
\sigma_{\delta\rho}\epsilon^{\omega\mu\delta\rho}$ and  contract the
Levi-Civita tensors. Using Gordon's identity again, the resulting
expression can be brought to the following form,
\begin{align}
  \mathcal{M}_{\mathcal{O}_2}
  &=
  -\frac{i\lambda^2e^2}{64\pi^2 m_\phi^2}
  \sum_q \langle N|  Q\, \bar{q}\gamma_\alpha q |N\rangle
  \nonumber\\&\qquad\times
  \left[ m_\chi\bar{u}(p_2) \sigma^{\alpha\beta} u(p_1)
  \frac{k_\beta}{k^2} \right.
  \nonumber\\&\qquad\qquad\qquad\qquad\left.
  + \frac i2 \bar{u}(p_2) \gamma^\alpha u(p_1) \right] \; .
\end{align}
Combining both operators, the total scattering amplitude
is
\begin{align}
  \mathcal{M}
  &=
  \sum_q
  \left[ {\mu_\chi e}
  \bar{u}(p_2) \sigma^{\alpha\beta} u(p_1)
  \frac{ik_\alpha}{k^2}
  \langle N| Q\, \bar{q}\gamma_\beta q |N\rangle
  \nonumber\right.
  \\&\qquad\qquad\left.
  + b_p \bar{u}(p_2) \gamma^\beta u(p_1)
  \langle N| Q\, \bar{q}\gamma_\beta q |N\rangle
  \right] \; ,
\end{align} where
we have defined
\begin{align}
  \mu_\chi
  &= \frac{\lambda^2 e
  m_\chi}{64 \pi^2 m_\phi^2}
  \\
  b_p
  &=
  \frac{\lambda^2 e^2}{64\pi^2
  m_\phi^2}
  (1+
  \frac23\log
  \left[\tfrac{m_{\ell}^2}{m_\phi^2}\right])
  \; ,
  \label{eqnbp}
\end{align}
and neglected the velocity-suppressed
contribution from $\mathcal{M}_{\mathcal{O}_1}$.

The first term in the amplitude corresponds to the magnetic dipole
moment of $\chi$ interacting with the nucleus, and the second term is
the familiar charge-charge interaction.  The dipole couples to both
the charge of the nucleus and its magnetic dipole moment. The
momentum-transfer dependence of each of these terms is different. The
dipole-charge interaction is enhanced at low-momentum transfers due
to the presence of the $k_\alpha/k^2$ factor. However, the coupling
to the dipole moment of the nucleon involves an additional power of
the momentum transfer $k$. Therefore the dipole-dipole interaction
has no such enhancement and exhibits the same recoil spectrum as the
charge-charge interaction up to form factors.

We show the three components of the scattering cross section:
charge-charge ($\sigma_{ZZ}$), dipole-charge ($\sigma_{DZ}$) and
dipole-dipole ($\sigma_{DD}$) \cite{Barger:2010gv,Chang:2010en}.  The
differential scattering cross sections with respect to the recoil
energy $E_r$, are given as follows,

\begin{align}
  \frac{d\sigma_{ZZ}}{dE_r} &= \frac{2 m_N}{4\pi  v^2} Z^2\, b_p^2\,
  F^2(E_r)
\end{align}
\begin{align}
  \frac{d\sigma_{DZ}}{dE_r}
  &=
  \frac{e^2 Z^2
  \mu_\chi^2}{4\pi E_r} \left[ 1-\frac{E_r}{v^2}
  \frac{m_\chi+2m_N}{2m_N m_\chi} \right]
  F^2(E_r)
\end{align}
\begin{align}
  \frac{d\sigma_{DD}}{dE_r}
  &=
  \frac{m_N\, \mu_{nuc}^2
  \,\mu_\chi^2}{\pi v^2}
  \left( \frac{S_{nuc}+1}{3S_{nuc}} \right)
  F_D^2(E_r) .
\end{align}
Here $m_N$ is the mass of the nucleus, $v$ is
the velocity of the dark matter particle.  $S_{nuc}$ is the spin of
the nucleus, $\mu_{nuc}$ is the magnetic dipole moment of the
nucleus, and $F_D(E_r)$ is the dipole moment form factor for the
nucleus.  The dipole-charge interaction is clearly enhanced at low
momentum transfer.

These results are only valid for the muon and the tau. However, in
the case of the electron, the only significant difference is that it
is the scale associated with the momentum transfer in the process
$|\vec{k}|$ that cuts off the logarithm in Eq.~\ref{eqnbp}, and not
the mass of the lepton $m_{\ell}$. In order to obtain
approximate limits for the case of electron flavored dark matter it
suffices to replace $m_{\ell}$ in Eq.~\ref{eqnbp} by $|\vec{k}|$. We
choose $|\vec{k}|$ = 10
MeV as a reference value.

\bibliographystyle{apsrev4-1}
\bibliography{fdm-ref}
\end{document}